\documentclass[11pt]{article}
\usepackage{jheppub}
\usepackage[space]{grffile}
\usepackage{epstopdf}
\usepackage{epsfig}
\usepackage{graphicx}   
\usepackage{float}
\usepackage{subfigure}  
\usepackage{verbatim}   
\usepackage{slashed}
\usepackage{amsmath,amssymb,graphicx}
\usepackage{amsfonts}
\usepackage{mathrsfs}
\usepackage{physics}
\usepackage{textcomp} 
\usepackage{gensymb} 
\usepackage{epsf,color}
\usepackage[dvipsnames]{xcolor}
\usepackage{url}
\usepackage{hyperref}
\definecolor{nicered}{rgb}{0.5,0.,0.}
\definecolor{nicegreen}{rgb}{0.,0.5,0.}
\definecolor{niceblue}{rgb}{0.,0.,0.5}
\hypersetup{colorlinks,citecolor=nicegreen,linkcolor=nicered,urlcolor=niceblue}
\usepackage{cancel}
\usepackage{framed}
\usepackage{array}
\usepackage{dcolumn}
\usepackage{cprotect}
\usepackage{multirow}
\usepackage{framed}
\usepackage{setspace}
\usepackage{enumitem}
\setlist{nolistsep}

\title{A CT18 global PDF fit at the leading order in QCD}

\author[a]{Mengshi Yan,}
\emailAdd{msyan@pku.edu.cn}
\affiliation[a]{Department of Physics and State Key Laboratory of Nuclear Physics and Technology, Peking University, Beijing 100871, China}

\author[b]{Tie-Jiun Hou,}
\emailAdd{tjhou@msu.edu}
\affiliation[b]{Center for Theory and Computation, National Tsing Hua University, Hsinchu 300, Taiwan}

\author[c]{Pavel Nadolsky,}
\emailAdd{nadolsky@mail.smu.edu}
\affiliation[c]{Department of Physics, Southern Methodist University, Dallas, TX 75275-0175, U.S.A.}

\author[d]{C.-P.  Yuan,}
\emailAdd{yuanch@msu.edu}
\affiliation[d]{Department of Physics and Astronomy,
	Michigan State University, East Lansing, MI 48824, U.S.A.}

\collaboration{CTEQ-TEA Collaboration}
\date{\today}

\abstract{
In this paper, we present a CT18 PDFs fitted with Leading-Order QCD perturbation theory. The CT18 LO PDFs is obtained within the general CT18 framework~\cite{Hou:2019efy}, along with two additional treatments being imposed to improve the quality of the fit. We take the $W$-boson charge asymmetry and inclusive single-top production at LHC as examples to illustrate the implication of the CT18 LO PDFs.
}

\preprint{\\ MSUHEP-22-013, SMU-HEP-22-05}

\begin{document}

\singlespacing

\maketitle

%

\section{Introduction}

Parton distribution functions (PDFs) describes the structure of hadrons as composed of (anti)quarks and gluons. PDFs is needed to make predictions for hard scattering processes in high-energy collisions.
Currently, with measurements at the Large Hadron Collider (LHC) becoming unprecedentedly precise, PDFs must be known at a high level of accuracy and precision. Such precise PDF parametrizations are provided by several groups~\cite{Hou:2019efy, Bailey:2020ooq, Ball:2021leu} by taking advantage of the availability of predictions at next-to-next-to-leading order (NNLO) in QCD coupling $\alpha_s$ for a large number of collider processes. Meanwhile, predictive power of leading order (LO) QCD theory is no longer sufficient for today's precise measurements. 
However, the need for LO PDFs still exists. 
Commonly used event generators, such as {\tt PYTHIA}, still rely on simulations of parton showers using LO splitting kernels~\cite{Sjostrand:2014zea},
though progress has been made in literature to implement parton showers at the next-to leading order (NLO) ~\cite{Nason:2004rx}.

The latest LO PDFs in the high-precision LHC era are MSHT20 LO~\cite{Bailey:2020ooq} and NNPDF4.0 LO~\cite{Ball:2021leu}. 
To improve the description of Drell-Yan (DY) processes,
a $K$-factor of $1 + \alpha_s C_F \pi/2$ has been adopted in LO PDFs fits of MSHT family~\cite{Martin:2009iq, Harland-Lang:2014zoa, Bailey:2020ooq}. The fit quality of MSHT20 LO is $\chi^2/N_{pt} = 2.58$, which is worse than their previous LO results and the MSHT20 PDFs beyond LO. The NNPDF4.0 LO is not able to fit experimental data well either, with $\chi^2/N_{pt} = 3.35$. Such bad qualities of fits for these LO results come from the inclusion of high precise LHC measurements, where NNLO corrections to theory predictions are already essential for a precise description.

The optimized CTEQ-TEA PDFs, CT09MCS, CT09MC1, and CT09MC2, dedicated for the use with event generator~\cite{Lai:2009ne} were obtained by generalizing the conventional QCD global analysis. The CT09MC candidate PDFs are constrained by not only the real experimental data sets, but also the NLO pseudo-data sets for representative LHC processes as joint input to the global fit, so that the description of the underlying event at the Tevatron and LHC is at a reasonably good level.
Specifically, in CT09MCS, the factorization scales in the LO matrix elements and the normalization for each pseudo-data sets are allowed to be varied to reach the best agreement with NLO pseudo-data.
In CT09MC1 and CT09MC2 PDFs, the total momentum sum rule, which reflects the conservation of total momentum carried by partons, is relaxed, and the factorization scales for pseudo-data sets are fixed. The normalization to each pseudo-data sets is still fitted as in the CT09MCS analysis. Two PDFs sets, CT09MC1 and CT09MC2 PDFs, are determined with 1-loop and 2-loop expression of $\alpha_s$ running respectively.
For the last LO result of CTEQ-TEA family, CT14 LO PDFs~\cite{Dulat:2015mca}, there are two PDFs sets provided. One is obtained with LO $\alpha_s$ evolution and $\alpha_s(M_Z) = 0.130$ for making up the insufficient PDFs evolution. The other PDFs set is obtained with NLO $\alpha_s$ evolution and $\alpha_s(M_Z) = 0.118$. The qualities of fits $\chi^2/N_{pt}$ are 1.99, 2.17 for CT14 LO with LO $\alpha_s$ evolution and with NLO $\alpha_s$ evolution, respectively. 

Coming to the CT18~\cite{Hou:2019efy} global analysis, a wide variety of precise LHC measurements, on top of the combine HERA I+II DIS~\cite{H1:2015ubc} data sets and CT14 data sets, are used in the determination of CT18 NLO and NNLO PDFs. We shall expect that a LO fit within the CT18 global analysis framework would suffer from the difficulty of describing high-precise LHC data without higher-order corrections, like the situation occurs in MSHT20 LO~\cite{Bailey:2020ooq} and NNPDF4.0 LO~\cite{Ball:2021leu}. 

In this work, we present the CT18 LO PDFs global fit within the general framework of CT18 NLO and NNLO studies~\cite{Hou:2019efy}. The experimental data sets and the settings of the fit are introduced in Sec.~\ref{sec:global_fitting}. Sec.~\ref{sec:resutls} describes the results of the fit and analysis on the impacts of the new LHC precision data. In Sec.~\ref{sec:pheno}, we study the implication of the CT18 LO PDFs by comparing some of the LHC observables, such as the $W$-boson charge asymmetry and the single-top production.
Sec.~\ref{sec:conclusion} contains our conclusion.

\section{Description of the CT18 LO global fit}
\label{sec:global_fitting}

In this section, we firstly introduce the experimental data sets as input for CT18 LO global fit. Then we impose two additional treatments to improve the results of fit from the limitation of LO perturbation theory.


\subsection{CT18 data sets}

The CT18 QCD global analysis~\cite{Hou:2019efy} is obtained by fitting from a wide range of LHC data sets with high precision, the combined HERA I+II DIS data sets, and data sets already included in CT14 global QCD analysis~\cite{Dulat:2015mca}, totally 40 data sets. The global fits are performed with NLO and NNLO QCD perturbation theories. Both NLO and NNLO fits can describe this large data set well in the measurement of $\chi^2/N_{pt}$ as shown in Tables~\ref{tab:quality_of_fit} and~\ref{tab:quality_of_fit_others}. In the following section~\ref{subsec:treatment}, we will show that a subset of CT18 data set, particularly consisting of high-precision LHC data sets in Run-II era, is difficult to fit with LO QCD theory.

For the high-precision LHC data sets included in the CT18 analyses, there are six data sets corresponding to $W$ and $Z$ vector boson production.
For the ATLAS measurements, they are the $\sqrt{s} = 7$ TeV $W$ and $Z$ combined cross-section measurement with 4.6 fb$^{-1}$ of integrated luminosity (ID=248)~\cite{ATLAS:2016nqi}, and $\sqrt{s} = 8$ TeV transverse momentum $p_T$ of lepton pairs distribution in the $Z/\gamma^*$ production with 20.3 fb$^{-1}$ of integrated luminosity (ID=253)~\cite{ATLAS:2015iiu}.
For the CMS measurement, the $\sqrt{s} = 8$ TeV muon charge asymmetry $A_{ch}$ for inclusive $W^{\pm}$ production with 18.8 fb$^{-1}$ of integrated luminosity (ID=249)~\cite{CMS:2016qqr} is included. 
For the LHCb measurements, the data sets included in CT18 are $\sqrt{s} = 7$ TeV $W$ and $Z$ forward rapidity cross-section distribution measurement with 1.0 fb$^{-1}$ of integrated luminosity (ID=245)~\cite{LHCb:2015okr}, $\sqrt{s} = 8$ TeV $Z \rightarrow e^+e^-$ forward rapidity cross-section distribution measurement with 2.0 fb$^{-1}$ of integrated luminosity (ID=246)~\cite{LHCb:2015kwa}, and $\sqrt{s} = 8$ TeV $W$ and $Z$ production cross-section distribution measurement with 2.0 fb$^{-1}$ of integrated luminosity (ID=250)~\cite{LHCb:2015mad},  respectively. 
The ATLAS $\sqrt{s} = 7$ TeV $W$ and $Z$ combined cross-section measurement (ID=248) data set~\cite{ATLAS:2016nqi} is not included in the nominal CT18, since this data set is observed to have tension with other data sets (Sec. II.C of~\cite{Hou:2019efy}). Alternative PDFs sets, CT18A and CT18Z, have been generated with the inclusion of ATLAS $\sqrt{s} = 7$ TeV $W$ and $Z$ data set.

In the analysis of CT18 LO, we start from the CT18 data sets, without the inclusion of the mentioned ATLAS $\sqrt{s} = 7$ TeV $W$ and $Z$ data set. Then, as will be shown in the next section, we shall exclude five data sets which cannot be correctly described at LO.

\subsection{Special treatments adapted in the CT18 LO fit}
\label{subsec:treatment}

In comparison to the PDF fits at NLO and NNLO, a naive fit at LO tends to be problematic.
Theory predictions at LO are less complicated than at higher orders, but they miss some contributions of quantum corrections that are especially vital for describing the most precise experimental data. 
For example, many NLO predictions of LHC cross-sections tend to be larger than predictions at LO in terms of magnitudes, see Figure 1 of Ref.~\cite{Lai:2009ne} for comparison between LO and NLO predictions of SM boson rapidity distributions at LHC.
On the other hand, shapes and magnitudes of PDFs are restricted by the momentum sum rule and the flavour number sum rules.
Consequently, PDFs determined at LO is known to have incorrect behaviour over wide range of $x$.
So predictions of spectrum with LO PDFs and LO matrix elements are unreliable also in terms of shape, for example see Figure 1 of Ref.~\cite{Lai:2009ne} for comparison between LO and NLO PDFs with LO matrix elements.
To resolve the difficultly in the determination of LO PDFs and the generation of LO predictions with LO PDFs, the conventional approach of PDFs determination is in need for extension.

In CT18 global analyses, several vector boson production data sets from LHC Run-II are included. The experimental uncertainty in these data sets is at a percent level and can strongly constrain PDFs at the electroweak scale. A LO fit with the inclusion of these data sets is difficult, and the best-fit PDFs fails to describe fitted data sets well. To illustrate this point, a PDF set named CT18 LOpert has been generated, where all theoretical predictions are computed at LO, and no other adjustments have been applied. In Tables~\ref{tab:quality_of_fit} and~\ref{tab:quality_of_fit_others}, the CT18 LOpert presents undesirable qualities of fit both totally and individually to specific data sets. In order to improve the fit at LO, in our final result, CT18 LO, we have applied the following two special treatments:

\begin{itemize}
\item From the CT18 data set, we exclude   	
ID 169 H1 $F_L$~\cite{H1:2015ubc},
ID 145 H1 bottom reduced cross-section~\cite{H1:2004esl}, ID 147 combined HERA charm production~\cite{H1:2012xnw},  ID 253 ATLAS 8 TeV $Z$ boson $p_T^{ll}$ distribution~\cite{ATLAS:2015iiu}, and 
ID 268 ATLAS 7 TeV $W$ and $Z$ bosons rapidity distribution plus $W$ charge asymmetry distribution~\cite{ATLAS:2011qdp}, since these data sets cannot be well described by the QCD theory at leading order. 
Furthermore, the ID248 ATLAS 7 TeV precision $W$ and $Z$ data~\cite{ATLAS:2016nqi} were not included in the nominal CT18 NLO and NNLO analysis, due to their tension with other datasets in the global fit~\cite{Hou:2019efy}. For comparison, alternative CT18A and CT18Z NLO and NNLO PDFs sets were generated~\cite{Hou:2019efy} with the inclusion of this data set.
After excluding all the above-mentioned data sets, the total number of remaining data points is 3547, which is 134 points less than the number of CT18 data set. 
\item Fot the rest of Drell-Yan data sets, for the inclusive production of either $W^\pm$ or $Z$ bosons, we adopt a $K$-factor $K(Q)$ as in Eq. \ref{Eq:K-fac} to partially make up the limitation of LO matrix elements, 

\end{itemize}
\begin{equation}
\label{Eq:K-fac}
    K(Q) = 1 + \frac{\alpha_s(Q)}{\pi}\frac{C_F\pi^2}{2}.
\end{equation}

\begin{table}[htbp]\small
\begin{center}  
\begin{tabular}{ll|llll} 
ID & Data set & CT18 LO & CT18 LOpert & CT18 NLO & CT18 NNLO \\ \hline \hline
145 & H1 $\sigma_r^b$~\cite{H1:2004esl} & 6.14$^*$ & 6.26 & 1.49 & 0.68 \\
147 & Combined HERA charm production~\cite{H1:2012xnw} & 21.14$^*$ & 11.54 & 0.80 & 1.24 \\
169 & H1 $F^L$~\cite{H1:2010fzx} & 17.15$^*$ & 17.15 & 0.77 & 1.89 \\
245 & LHCb 7 TeV W/Z rap.~\cite{LHCb:2015okr} & 5.85 & 8.36 & 2.29 & 1.63 \\
246 & LHCb 8 TeV Z$\rightarrow$ee rap.~\cite{LHCb:2015kwa} & 5.84 & 11.06 & 2.09 & 1.00 \\
249 & CMS 8 TeV W A$_{ch}$~\cite{CMS:2016qqr} & 2.17 & 9.14 & 0.60 & 1.03 \\
250 & LHCb 8 TeV W/Z rap.~\cite{LHCb:2015mad} & 10.59 & 13.61 & 3.36 & 2.17 \\
253 & ATLAS 8 TeV Z $p_{T}^{ll}$~\cite{ATLAS:2015iiu} & 19.21$^*$ & 19.20 & 2.06 & 1.12 \\ \hline
 & total & 1.60 & 2.15 & 1.17 & 1.17
\end{tabular}  
\end{center}  
\caption{\label{tab:quality_of_fit}
The qualities of fits for selected data sets, whose $\chi^2/N_{\text{pt}}$ of CT18 LOpert are larger than 6.0. The values with an asterisk are $\chi^2/N_{\text{pt}}$ of the data sets which are not included in the corresponding fit.
}
\end{table}

\begin{table}[htbp]\small
\begin{center}  
\begin{tabular}{ll|llll}
ID & Data set & CT18 LO & CT18 LOpert & CT18 NLO & CT18 NNLO \\ \hline \hline
160 & HERAI+II 1 fb$^{-1}$ H1 and ZEUS NC & 1.72 & 1.64 & 1.22 & 1.26 \\
 & and CC e$^{\pm}$p reduced cross sec. comb.~\cite{H1:2015ubc} & & & & \\
101 & BCDMS $F_2^p$~\cite{BCDMS:1989ggw} & 1.16 & 1.26 & 1.08 & 1.11 \\
102 & BCDMS $F_2^d$~\cite{BCDMS:1989qop} & 1.29 & 1.64 & 1.13 & 1.12 \\
104 & NMC $F_2^d/F_2^p$~\cite{NewMuon:1996fwh} & 1.17 & 1.43 & 0.97 & 1.02 \\
108 & CDHSW $F_2^p$~\cite{Berge:1989hr} & 1.09 & 1.52 & 0.90 & 1.01 \\
109 & CDHSW $F_3^p$~\cite{Berge:1989hr} & 1.22 & 1.42 & 0.82 & 0.90 \\
110 & CCFR $F_2^p$~\cite{CCFRNuTeV:2000qwc} & 1.78 & 2.55 & 1.15 & 1.14 \\
111 & CCFR $xF_3^p$~\cite{Seligman:1997mc} & 0.53 & 0.80 & 0.41 & 0.39 \\
124 & NuTeV $\nu\mu\mu$ SIDIS~\cite{Mason:2006qa} & 1.02 & 2.11 & 0.52 & 0.49 \\
125 & NuTeV $\bar{\nu}\mu\mu$ SIDIS~\cite{Mason:2006qa} & 1.73 & 2.46 & 1.03 & 1.17 \\
126 & CCFR $\nu\mu\mu$ SIDIS~\cite{NuTeV:2001dfo} & 0.60 & 1.43 & 0.79 & 0.75 \\
127 & CCFR $\bar{\nu}\mu\mu$ SIDIS~\cite{NuTeV:2001dfo} & 0.58 & 1.15 & 0.54 & 0.52 \\ \hline
504 & CDF Run-2 inclusive jet production~\cite{CDF:2008hmn} & 1.54 & 1.46 & 1.49 & 1.70 \\
514 & D$\emptyset$ Run-2 inclusive jet production~\cite{D0:2008nou} & 1.22 & 1.38 & 1.07 & 1.03 \\
542 & CMS 7 TeV 5 fb$^{-1}$ single incl. jet & 1.41 & 1.56 & 1.23 & 1.23 \\
 & cross sec., $R$ = 0.7 (extended in $y$)~\cite{CMS:2014nvq} & & & & \\ 
544 & ATLAS 7 TeV 4.5 fb$^{-1}$ single incl. jet & 1.50 & 1.58 & 1.40 & 1.45 \\
 & cross sec., $R$ = 0.6~\cite{ATLAS:2014riz} & & & & \\
545 & CMS 8 TeV 19.7 fb$^{-1}$ single incl. jet & 1.65 & 1.78 & 1.10 & 1.14 \\
 & cross sec., R = 0.7, (extended in $y$)~\cite{CMS:2016lna} & & & & \\
573 & CMS 8 TeV 19.7 fb$^{-1}$ $t\bar{t}$ norm. double-diff.  & 2.05 & 2.38 & 1.56 & 1.18 \\
 & top $p_T$ and $y$ cross sec. single~\cite{CMS:2017iqf} & & & & \\
580 & ATLAS 8 TeV 20.3 fb$^{-1}$ $t\bar{t}$ $p_T^t$ and $m_{t\bar{t}}$ & 2.41 & 3.38 & 1.29 & 0.63 \\
 & absolute spectrum~\cite{ATLAS:2015lsn} & & & & \\
 \hline
 & total & 1.60 & 2.15 & 1.17 & 1.17
\end{tabular}  
\end{center}  
\caption{\label{tab:quality_of_fit_others}
Similar to Table~\ref{tab:quality_of_fit}, qualities of fits $\chi^2/N_{\text{pt}}$ for DIS and jet data sets in the CT18 data set are compared. Totally, the number of data points in CT18 LO is 3547, which is reduced by the first special treatment from the the number of data points in CT18 LOpert, NLO and NNLO, 3861.
}
\end{table}

To elaborate on the first treatment, we first note that 
the longitudinal structure function $F_L = F_2 - 2xF_1$ at leading order $\mathcal{O}(\alpha_s^0)$ respects the Callan-Gross relation~\cite{Callan:1969uq},
\begin{equation}
    F_2 - 2xF_1 = 0.
\end{equation}
Beyond LO, the gluon emission would give arise of a non-vanishing $F_L$, and the Callan-Gross relation is violated accordingly.
In CT18 data set, the ID 169 H1 $F_L$ data~\cite{H1:2010fzx} measures the longitudinal structure function by $e^{\pm}p$ collision. With the reason just said, a LO PDF fit can never be able to describe this data. Hence we exclude this data set from the LO fit.
The ID 145 H1 bottom reduced cross-section~\cite{H1:2004esl} and the ID 147 Combined HERA charm production~\cite{H1:2012xnw} data measure the inclusive bottom and charm production rates, respectively, from deep inelastic $ep$ scattering,  
which are sensitive to higher order QCD corrections due to the non-vanishing mass of heavy partons of the proton. Detailed discussions can be found in Refs.~\cite{Aivazis:1993pi,Aivazis:1993kh,Kramer:2000hn,Tung:2001mv,Guzzi:2011ew}.
Therefore, we also exclude these two data sets.
Moreover, a LO QCD calculation, at $\mathcal{O}(\alpha_s)$, cannot describe well the ID 253 ATLAS 8 TeV $Z$ boson $p_T^{ll}$ distribution~\cite{ATLAS:2015iiu}, because of the presence of large logarithm  $\ln(M_Z/p_T^{ll})$. Likewise, a QCD calculation, at $\mathcal{O}(\alpha_s)$,  cannot describe well the inclusive $W$ or $Z$ productions, but with asymmetric kinematic cuts applied to the two decay leptons of the vector boson, such as  
the ID 268 ATLAS 7 TeV $W$ charge lepton rapidity asymmetry measurement ~\cite{ATLAS:2011qdp}. Needless to say that at  $\mathcal{O}(\alpha_s^0)$, the $p_T$ distribution of the Drell-Yan pair produced at the LHC is a delta function with peak at zero, and the two decay leptons must have the same transverse momenta, as predicted by the parton model with longitudinal PDFs. Hence, these data sets are also excluded in our LO fits.  

The importance of the second treatment will be discussed in Sec.~\ref{subsec:LHC_run2}. In LO PDFs studies of the MSHT family~\cite{Martin:2009iq, Harland-Lang:2014zoa, Bailey:2020ooq}, this $K$-factor, Eq. \ref{Eq:K-fac}, was also adopted for helping describing vector boson production data. 
Apart from these two special treatments, there is no other treatment been applied to CT18 LO, as we wish to keep the balance between a LO PDFs fit and a good fit quality.
As for the strong coupling $\alpha_s$ as input, it is expected that in a LO PDFs analysis the best-fit would prefer a larger value of $\alpha_s$. We decide to fix the strong coupling at $Z$-boson mass scale to be $\alpha_s(m_Z) = 0.135$ for CT18 LO. The  $\alpha_s$ dependence of CT18 LO fit will be discussed in Sec.~\ref{subsec:as}.

\section{Results}
\label{sec:resutls}

In this section, we present results of the CT18 LO PDFs fit, which is obtained based on the CT18 framework but with two additional treatments as defined in Sec.~\ref{subsec:treatment}.
Along with the fit quality, the presentation of PDF configuration and various PDF moments, the impact of Drell-Yan data from LHC precision measurements and $\alpha_s$ dependence of the fit will also be discussed.


\subsection{Quality of the fit}

Goodness-of-fit figures, $\chi^2/N_{pt}$, for selected data sets are summarised in Tables~\ref{tab:quality_of_fit},~\ref{tab:quality_of_fit_others}. 
The overall $\chi^2/N_{pt}$ of CT18 LO is 1.60, significantly improved from 2.15, the total $\chi^2/N_{pt}$ of CT18 LOpert fit. In total, the fit to the CT18 LO data set is clearly enhanced by two special treatments introduced in Sec.~\ref{subsec:treatment}, though it is still much worse than the CT18 NLO and NNLO fits.

For individual data sets, the majority of them receives a smaller $\chi^2/N_{pt}$ in CT18 LO, comparing to in CT18 LOpert. 
With the help of the second special treatment, the Drell-Yan $K$-factor, Eq. \eqref{Eq:K-fac}, the high-precision LHC $W$ and $Z$ bosons production data sets in Table~\ref{tab:quality_of_fit} obtain a better $\chi^2/N_{pt}$ in CT18 LO than in CT18 LOpert. But the fits to them by no means are good.
For DIS and jet data sets shown in Table~\ref{tab:quality_of_fit_others}, fits to these data sets are mostly improved from in CT18 LOpert, but again it is difficult to obtain a good fit with $\chi^2/N_{pt} \sim \mathcal{O}(1)$ at this order.
In CT18 LO the ID160 HERA I+II combined reduced cross-section~\cite{H1:2015ubc} has a slightly larger $\chi^2/N_{pt}$ than in CT18 LOpert. This increase only comes from several data points with very low energy scale $Q = 2.121$ GeV in the neutral current channel of $e^+p$ collision, where the correlated systematic errors pull the central values of data points away from theory prediction. In the CTEQ-TEA program, the best-fit value of $\chi^2$ is the combination of the best-fit $\chi^2$ to the shifted data and the contribution from the optimal nuisance parameters~\cite{Hou:2019efy, Pumplin:2002vw}.
The optimal nuisance parameters for these low energy neutral current HERA data points thus have very large values, suggesting that there is a noticeable systematical bias in the CT18 LO fit to these data points.
The fits to the rest of HERA data points for CT18 LO and CT18 LOpert are in general comparable. A similar phenomenon, that one data point being pulled far away by correlated systematic errors results in an increase in $\chi^2/N_{pt}$ from CT18 LOpert to CT18 LO, also happens to ID504 CDF Run-2 inclusive jet production~\cite{CDF:2008hmn}.

The equivalent information of the agreement to experimental data can be provided by the effective Gaussian variables $S = \sqrt{2 \chi^2} - \sqrt{2 N_{pt} - 1}$~\cite{Lai:2010vv}, whose distribution theoretically approaches $\mathcal{N}(0, 1)$ for large $N_{pt}$.
In Table~\ref{tab:spartyness}, we summarize the effective Gaussian variables of CT18 LO and CT18 LOpert for selected data sets in Tables~\ref{tab:quality_of_fit},~\ref{tab:quality_of_fit_others}.
We notice that there is a plenty of data sets in CT18 LO and CT18 LOpert having $|S| > 1$, so that totally the distribution of the effective Gaussian variable in CT18 LO deviates significantly from $\mathcal{N}(0, 1)$, which is expected in a good fit. 
The distribution of the effective Gaussian variables $S$, along with the distribution of $\chi^2/N_{pt}$, indicates that the experiments are strongly underfitted in CT18 LO, although two special treatments introduced in Sec.~\ref{subsec:treatment} can help improving the quality of the fit.

\begin{table}[htbp]\small
\begin{center}  
\begin{tabular}{ll|llll}
ID & Data set & CT18 LO & CT18 LOpert & CT18 NLO & CT18 NNLO \\ \hline \hline
145 & H1 $\sigma_r^b$~\cite{H1:2004esl} & 5.93$^*$ & 6.02 & -0.32 & -0.65 \\
147 & Combined HERA charm production~\cite{H1:2012xnw} & 30.13$^*$ & 20.06 & 2.15 & 1.15 \\
169 & H1 $F^L$~\cite{H1:2010fzx} & 11.54$^*$ & 11.54 & -0.36 & 1.65 \\
245 & LHCb 7 TeV W/Z rap.~\cite{LHCb:2015okr} & 10.23 & 13.41 & 3.99 & 2.24 \\
246 & LHCb 8 TeV Z$\rightarrow$ee rap.~\cite{LHCb:2015kwa} & 7.38 & 11.76 & 2.55 & 1.42 \\
249 & CMS 8 TeV W A$_{ch}$~\cite{CMS:2016qqr} & 2.22 & 8.32 & -0.95 & 0.22 \\
250 & LHCb 8 TeV W/Z rap.~\cite{LHCb:2015mad} & 16.09 & 19.10 & 6.34 & 3.73 \\
253 & ATLAS 8 TeV Z $p_{T}^{ll}$~\cite{ATLAS:2015iiu} & 21.45$^*$ & 21.44 & 3.10 & 0.51 \\ \hline
160 & HERAI+II 1 fb$^{-1}$ H1 and ZEUS NC & 14.06 & 12.78 & 4.92 & 5.66 \\
 & and CC e$^{\pm}$p reduced cross sec. comb.~\cite{H1:2015ubc} & & & & \\
101 & BCDMS $F_2^p$~\cite{BCDMS:1989ggw} & 2.05 & 3.19 & 1.05 & 1.39 \\
102 & BCDMS $F_2^d$~\cite{BCDMS:1989qop} & 3.01 & 6.02 & 1.40 & 1.34 \\
104 & NMC $F_2^d/F_2^p$~\cite{NewMuon:1996fwh} & 1.32 & 3.00 & -0.22 & 0.21 \\
108 & CDHSW $F_2^p$~\cite{Berge:1989hr} & 0.64 & 3.00 & -0.64 & 0.10 \\
109 & CDHSW $F_3^p$~\cite{Berge:1989hr} & 1.46 & 2.65 & -1.31 & -0.66 \\
110 & CCFR $F_2^p$~\cite{CCFRNuTeV:2000qwc} & 3.80 & 6.57 & 0.88 & 0.86 \\
111 & CCFR $xF_3^p$~\cite{Seligman:1997mc} & -3.67 & -1.36 & -5.09 & -5.24 \\
124 & NuTeV $\nu\mu\mu$ SIDIS~\cite{Mason:2006qa} & 0.14 & 3.80 & -2.51 & -2.72 \\
125 & NuTeV $\bar{\nu}\mu\mu$ SIDIS~\cite{Mason:2006qa} & 2.53 & 4.39 & 0.19 & 0.72 \\
126 & CCFR $\nu\mu\mu$ SIDIS~\cite{NuTeV:2001dfo} & -1.98 & 1.77 & -1.27 & -1.17 \\
127 & CCFR $\bar{\nu}\mu\mu$ SIDIS~\cite{NuTeV:2001dfo} & -2.08 & 0.70 & -2.34 & -2.49 \\ \hline
504 & CDF Run-2 inclusive jet production~\cite{CDF:2008hmn} & 2.85 & 2.46 & 2.60 & 3.54 \\
514 & D$\emptyset$ Run-2 inclusive jet production~\cite{D0:2008nou} & 1.59 & 2.55 & 0.55 & 0.30 \\
542 & CMS 7 TeV 5 fb$^{-1}$ single incl. jet & 3.28 & 4.33 & 1.97 & 1.96 \\
 & cross sec., $R$ = 0.7 (extended in $y$)~\cite{CMS:2014nvq} & & & & \\ 
544 & ATLAS 7 TeV 4.5 fb$^{-1}$ single incl. jet & 3.65 & 4.19 & 3.01 & 3.34 \\
 & cross sec., $R$ = 0.6~\cite{ATLAS:2014riz} & & & & \\
545 & CMS 8 TeV 19.7 fb$^{-1}$ single incl. jet & 5.28 & 6.16 & 0.96 & 1.30 \\
 & cross sec., R = 0.7, (extended in $y$)~\cite{CMS:2016lna} & & & & \\
573 & CMS 8 TeV 19.7 fb$^{-1}$ $t\bar{t}$ norm. double-diff.  & 2.41 & 2.97 & 1.48 & 0.60 \\
 & top $p_T$ and $y$ cross sec. single~\cite{CMS:2017iqf} & & & & \\
580 & ATLAS 8 TeV 20.3 fb$^{-1}$ $t\bar{t}$ $p_T^t$ and $m_{t\bar{t}}$ & 2.94 & 4.28 & 0.86 & -1.07 \\
 & absolute spectrum~\cite{ATLAS:2015lsn} & & & & \\ \hline
\end{tabular}  
\end{center}  
\caption{\label{tab:spartyness}
The effective Gaussian variables $S$ for selected data sets in Tables~\ref{tab:quality_of_fit},~\ref{tab:quality_of_fit_others}.
}
\end{table}

\subsection{Functional dependence and moments of PDFs}

\begin{figure}[htbp]
\centering
\hspace{0.01\columnwidth} 
\centering \subfigure[]{ \label{fig:PDF_configuration_1.3GeV:a} 
\includegraphics[width=0.475\columnwidth]{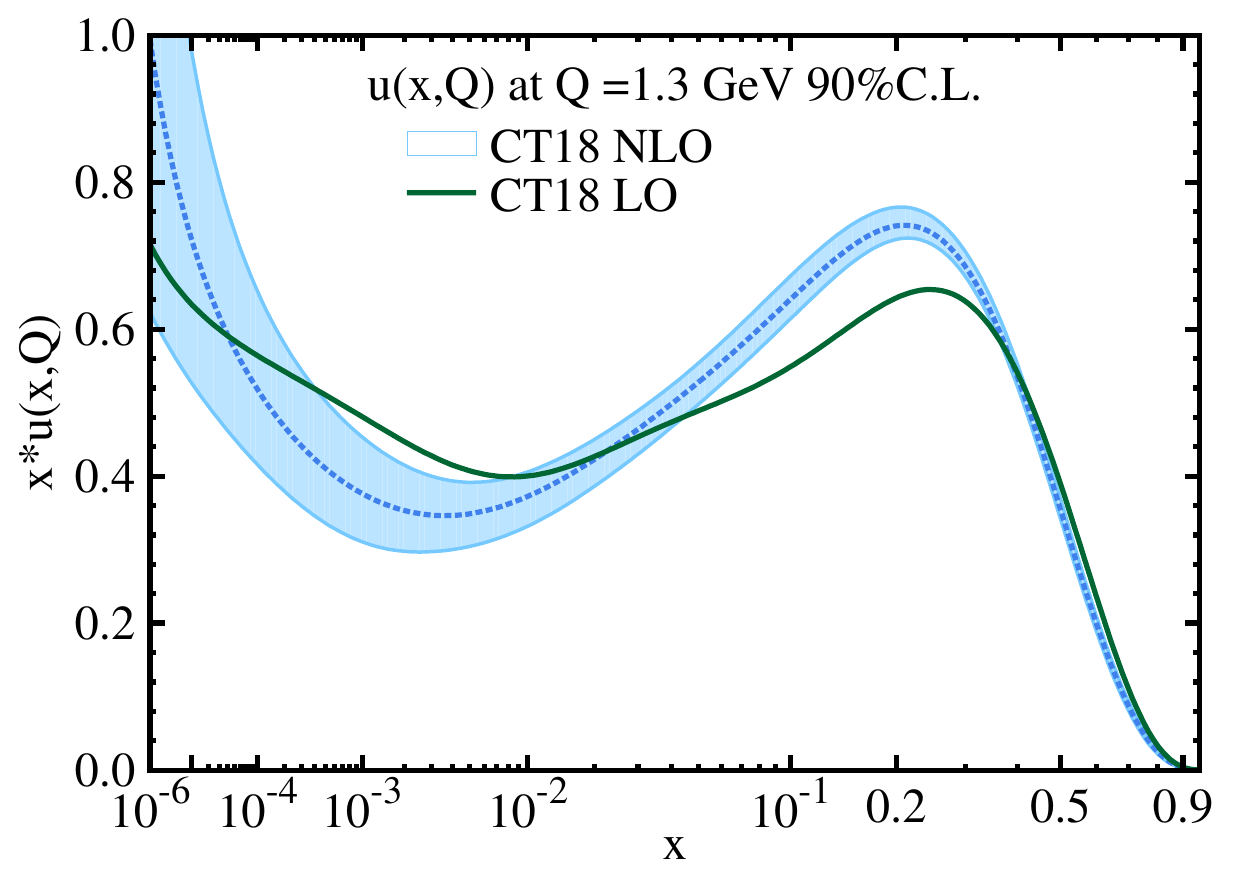}}  
\centering \subfigure[]{ \label{fig:PDF_configuration_1.3GeV:b} 
\includegraphics[width=0.475\columnwidth]{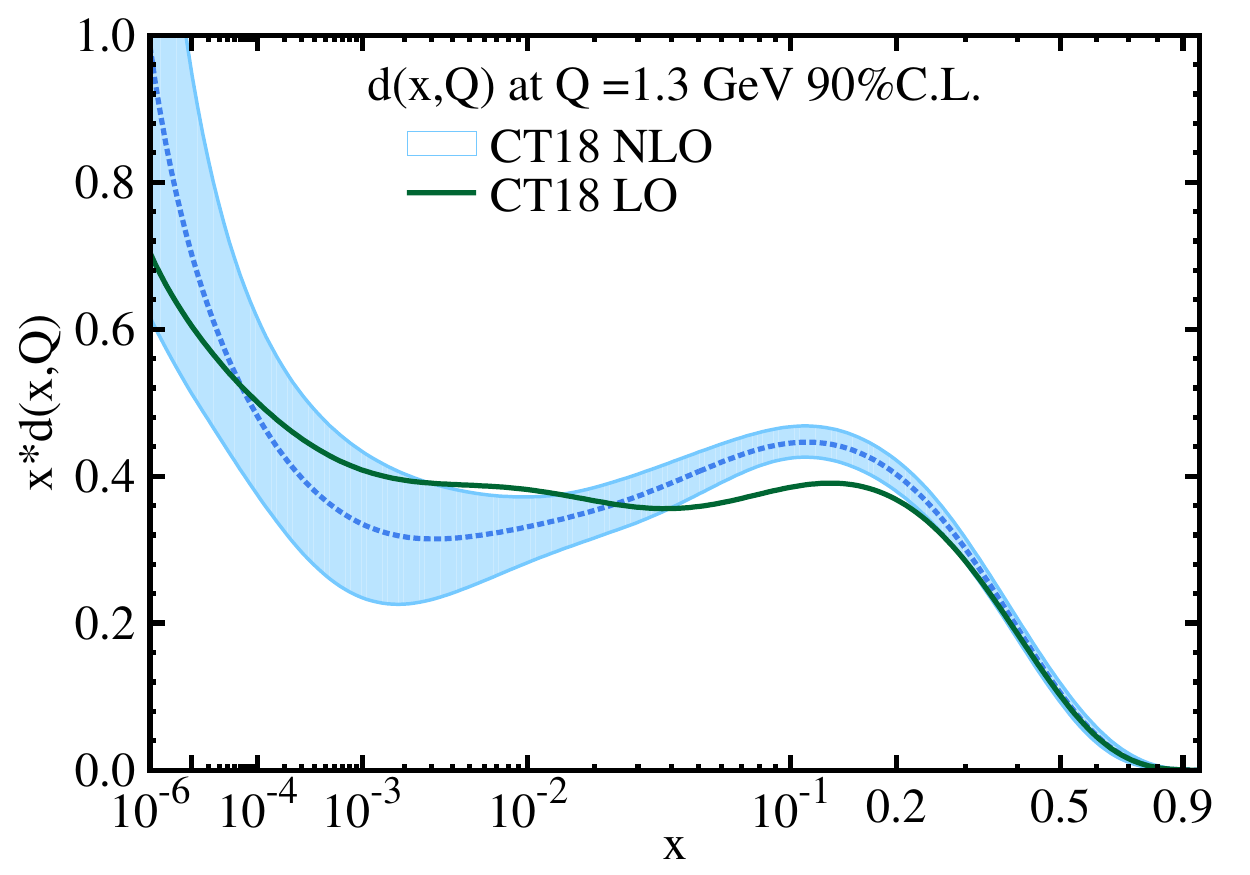}} 
\centering \subfigure[]{ \label{fig:PDF_configuration_1.3GeV:c} 
\includegraphics[width=0.475\columnwidth]{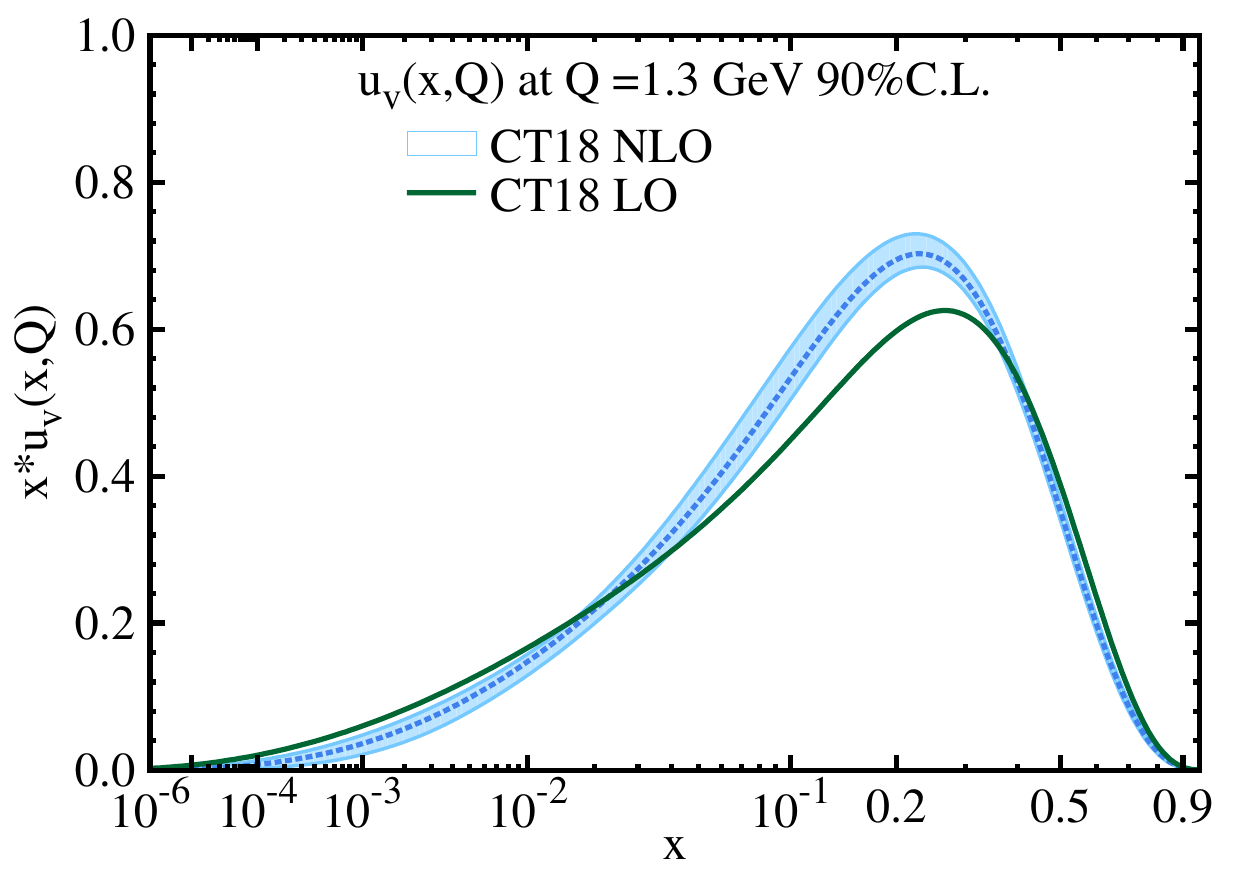}}  
\centering \subfigure[]{ \label{fig:PDF_configuration_1.3GeV:d} 
\includegraphics[width=0.475\columnwidth]{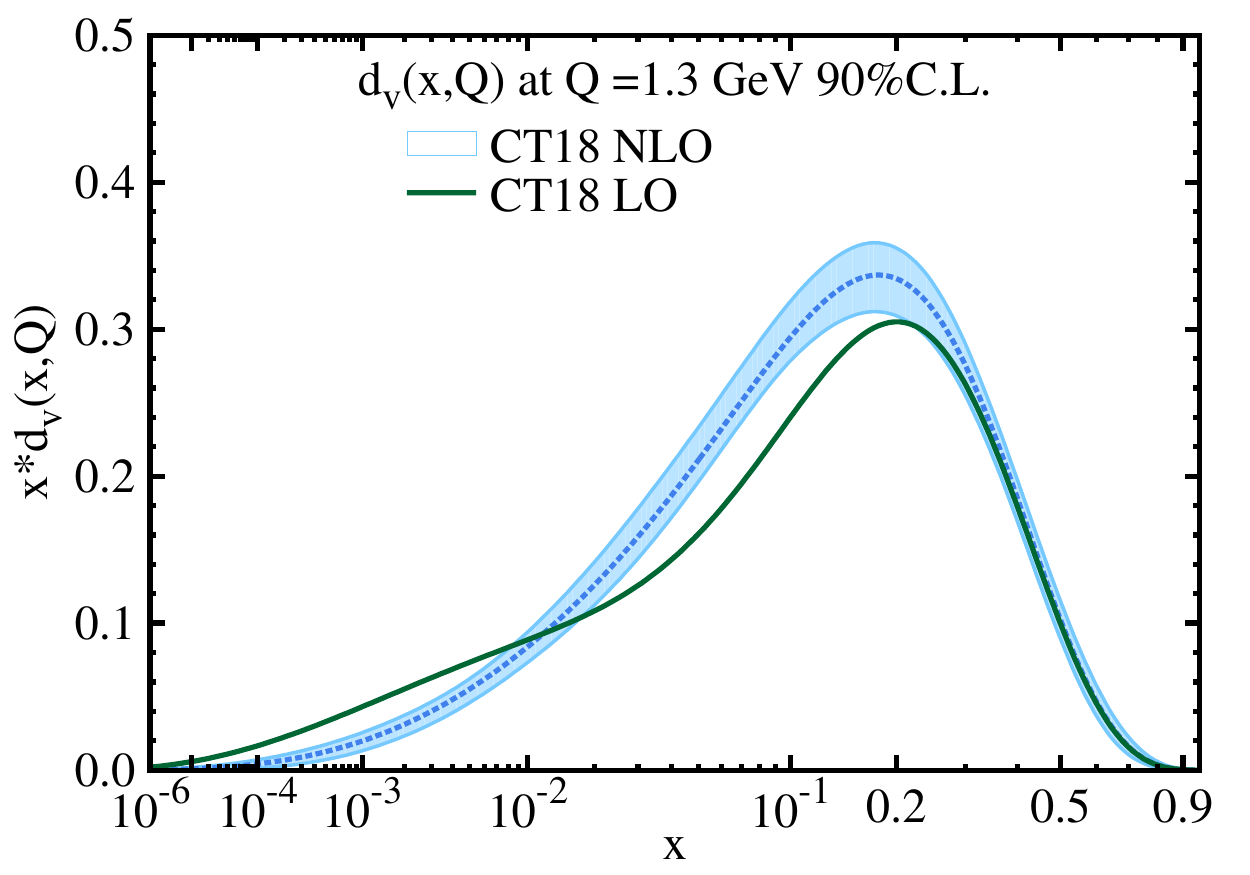}} 
\centering \subfigure[]{ \label{fig:PDF_configuration_1.3GeV:e} 
\includegraphics[width=0.475\columnwidth]{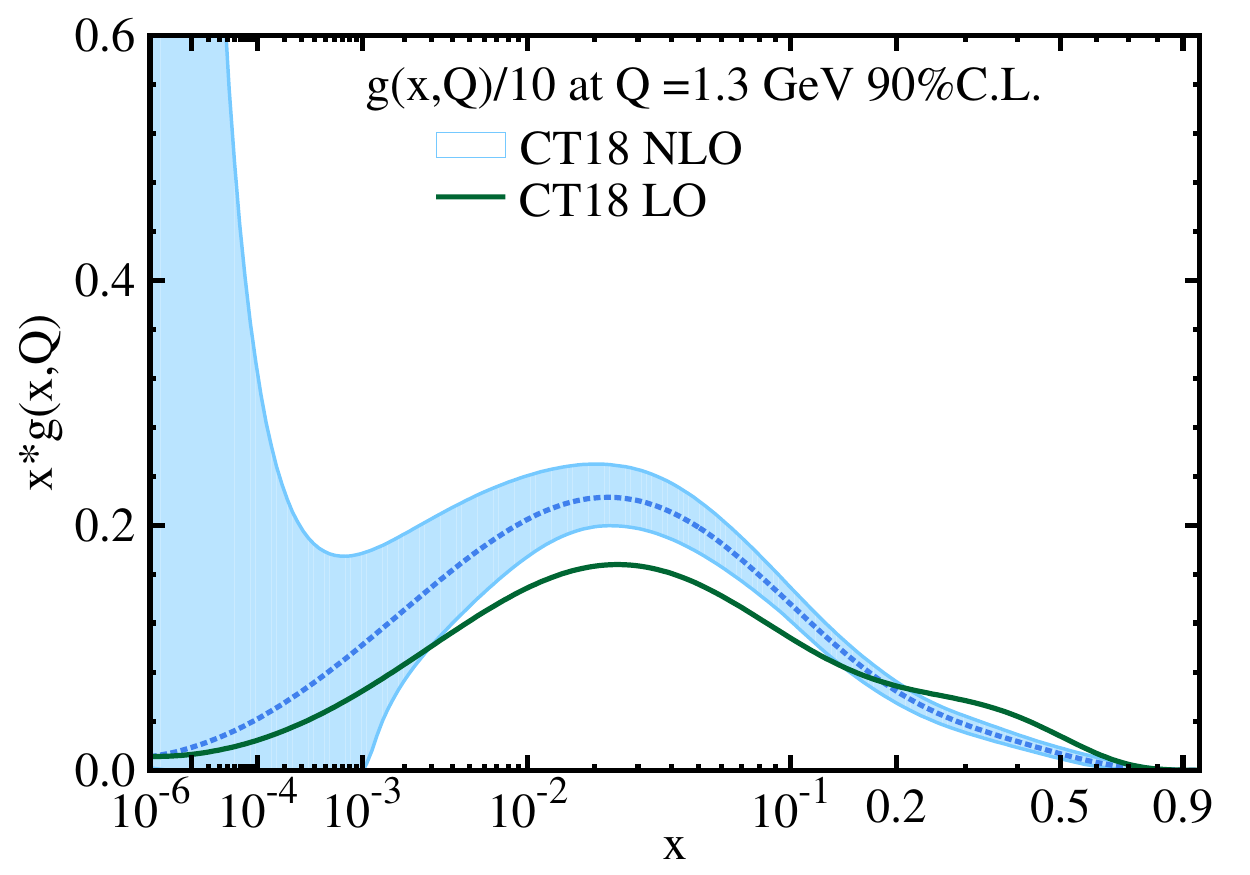}}  
\centering \subfigure[]{ \label{fig:PDF_configuration_1.3GeV:f} 
\includegraphics[width=0.475\columnwidth]{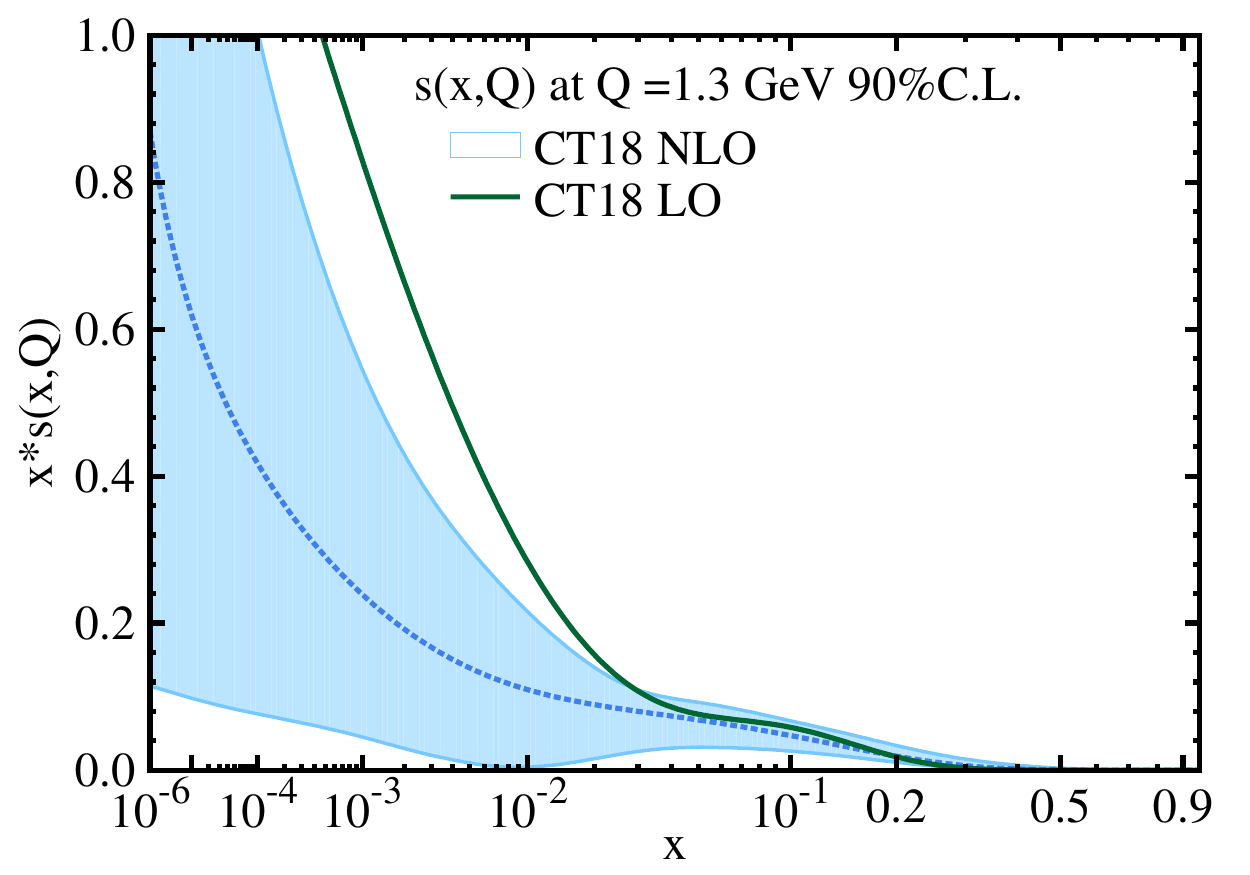}} 
\caption{The CT18 LO PDFs at initial scale $Q_0 = 1.3$ GeV compared with CT18 NLO PDFs.
}
\label{fig:PDF_configuration_1.3GeV} 
\end{figure}

Figure~\ref{fig:PDF_configuration_1.3GeV} compares CT18 LO PDFs and CT18 NLO PDFs at the inital scale $Q_0 = 1.3$ GeV.
In comparison to CT18 NLO, the CT18 LO exhibits a different configuration due to the significant difference between the LO and NLO QCD perturbation theory. In Figs.~\ref{fig:PDF_configuration_1.3GeV:a} and~\ref{fig:PDF_configuration_1.3GeV:b} for $u(x)$ and $d(x)$, the CT18 LO in the small-$x$ and large-$x$ limits is consistent with CT18 NLO. As shown in Figs.~\ref{fig:PDF_configuration_1.3GeV:c} and~\ref{fig:PDF_configuration_1.3GeV:d},  the bumps in valence quarks around $0.1 < x < 0.3$ are reduced so that CT18 LO is outside of the CT18 NLO error band. Meanwhile, the $u_v$ and $d_v$ distributions are enhanced in the region $x < 0.01$ due to the valence number sum rules.
The CT18 LO gluon PDF exhibits deviations from CT18 NLO in the ranges $3\times10^{-3} < x < 0.1$ and $x > 0.2$, cf.  Fig. ~\ref{fig:PDF_configuration_1.3GeV:e}. Enhancement of the gluon PDF in the large-$x$ region 
is needed at LO in order to compensate for the missing higher order contribution to the Wilson coefficients of a number of scattering processes, such as high $p_T$ jet production at Tevatron and LHC and the precision DIS data at HERA, as required by a  consistent NLO (or NNLO) theory calculation for  describing the existing data.
Fig.~\ref{fig:PDF_configuration_1.3GeV:f} shows that the CT18 LO strange PDF has a good agreement with CT18 NLO for $x > 0.02$. In the range $x < 0.02$, the CT18 LO strange PDF shows a larger magnitude than the CT18 NLO. 

\begin{figure}[htbp]
\centering
\hspace{0.01\columnwidth} 
\centering \subfigure[]{ \label{fig:PDF_configuration_100GeV:a} 
\includegraphics[width=0.475\columnwidth]{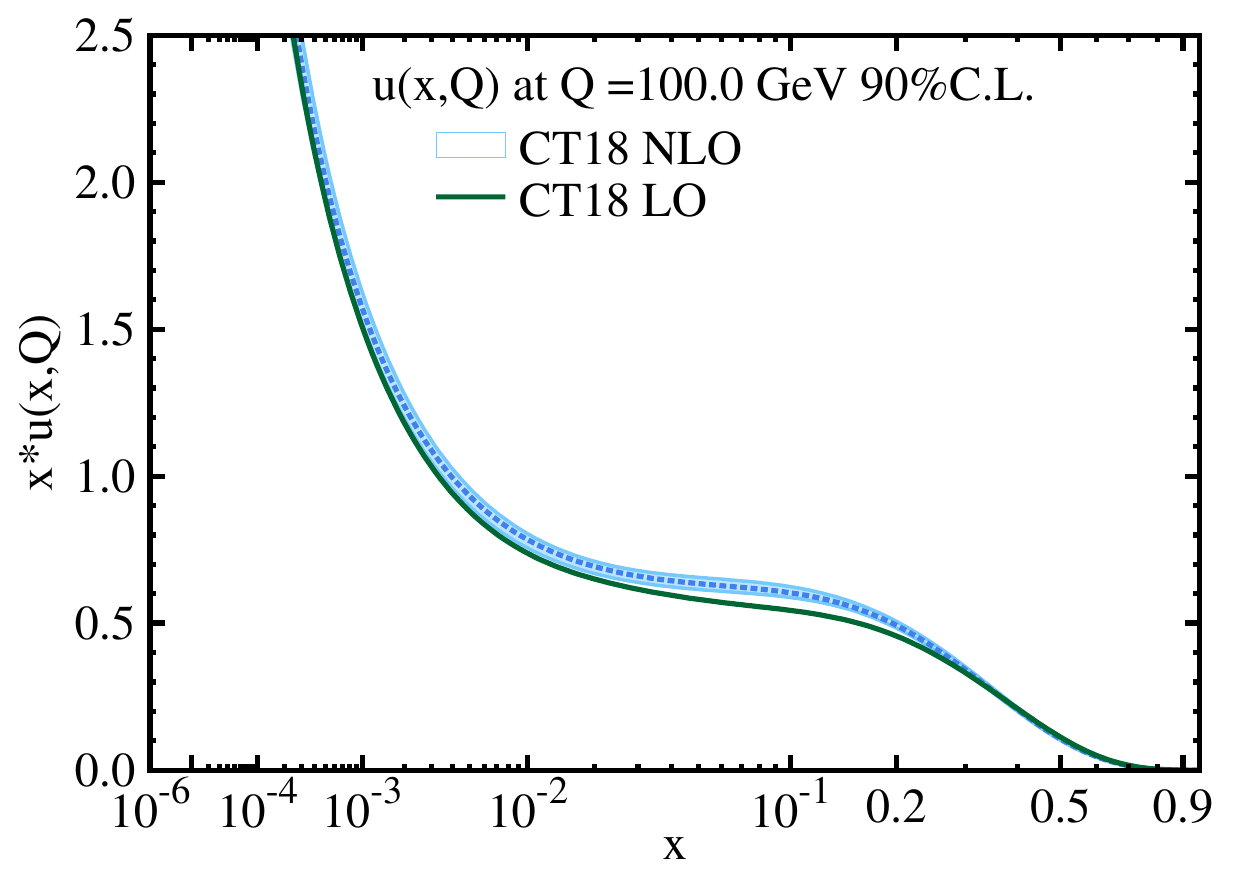}}  
\centering \subfigure[]{ \label{fig:PDF_configuration_100GeV:b} 
\includegraphics[width=0.475\columnwidth]{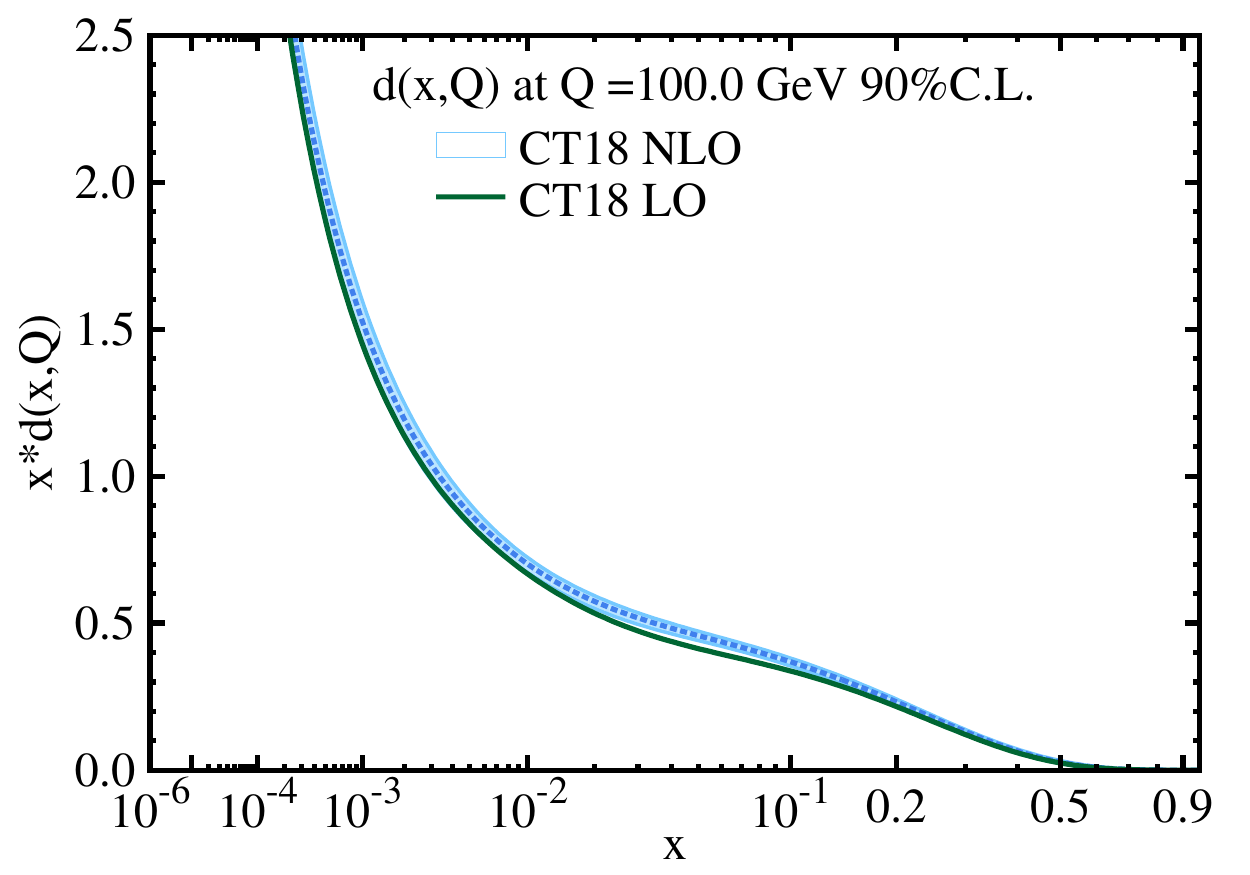}} 
\centering \subfigure[]{ \label{fig:PDF_configuration_100GeV:c} 
\includegraphics[width=0.475\columnwidth]{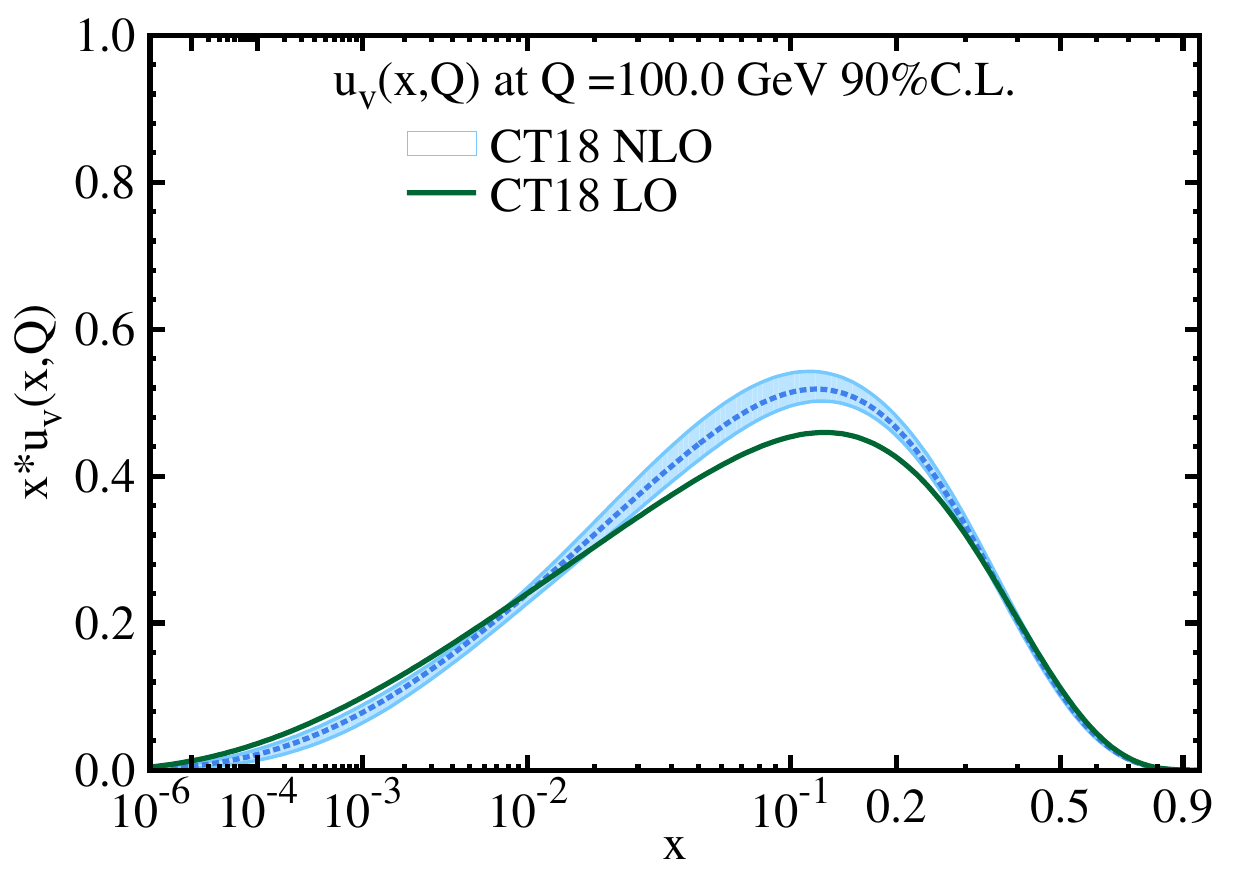}}  
\centering \subfigure[]{ \label{fig:PDF_configuration_100GeV:d} 
\includegraphics[width=0.475\columnwidth]{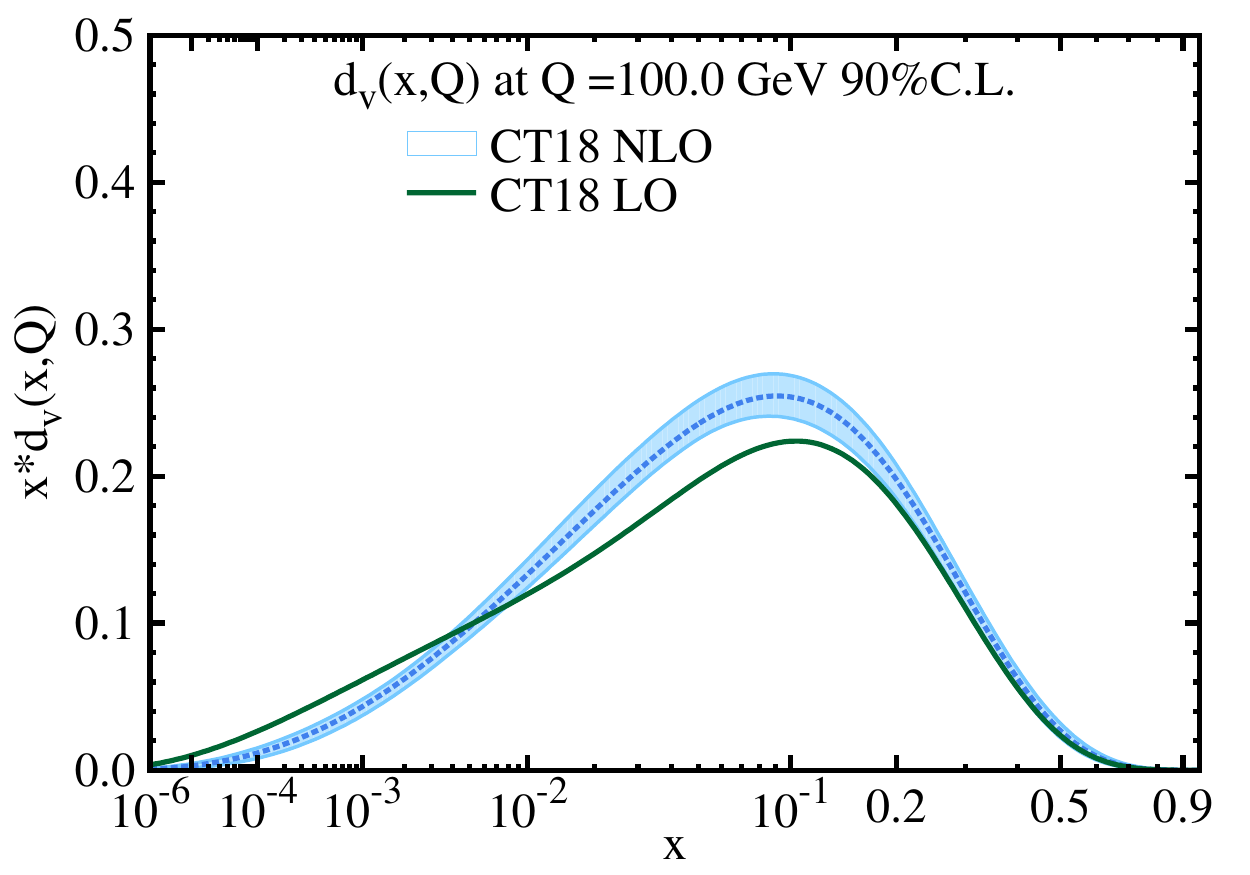}} 
\centering \subfigure[]{ \label{fig:PDF_configuration_100GeV:e} 
\includegraphics[width=0.475\columnwidth]{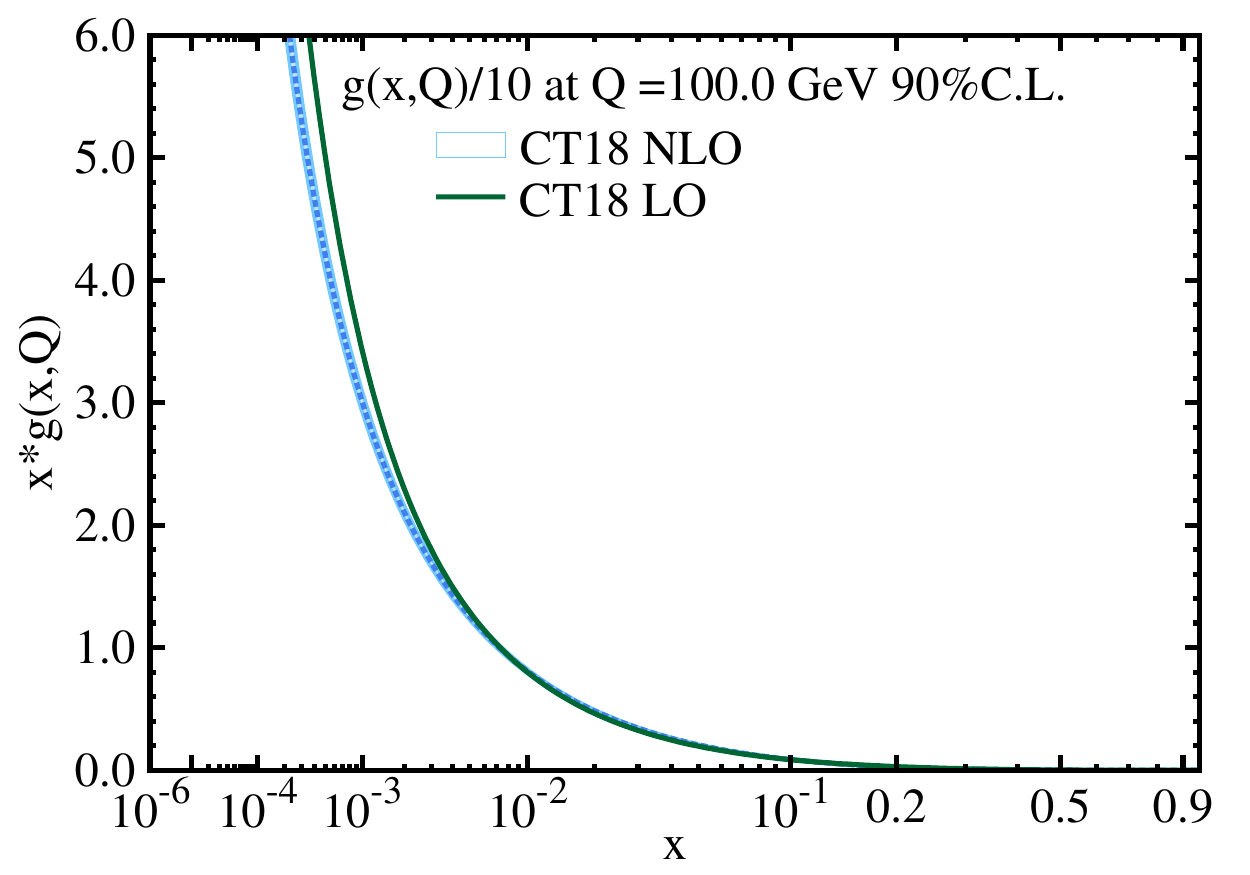}}  
\centering \subfigure[]{ \label{fig:PDF_configuration_100GeV:f} 
\includegraphics[width=0.475\columnwidth]{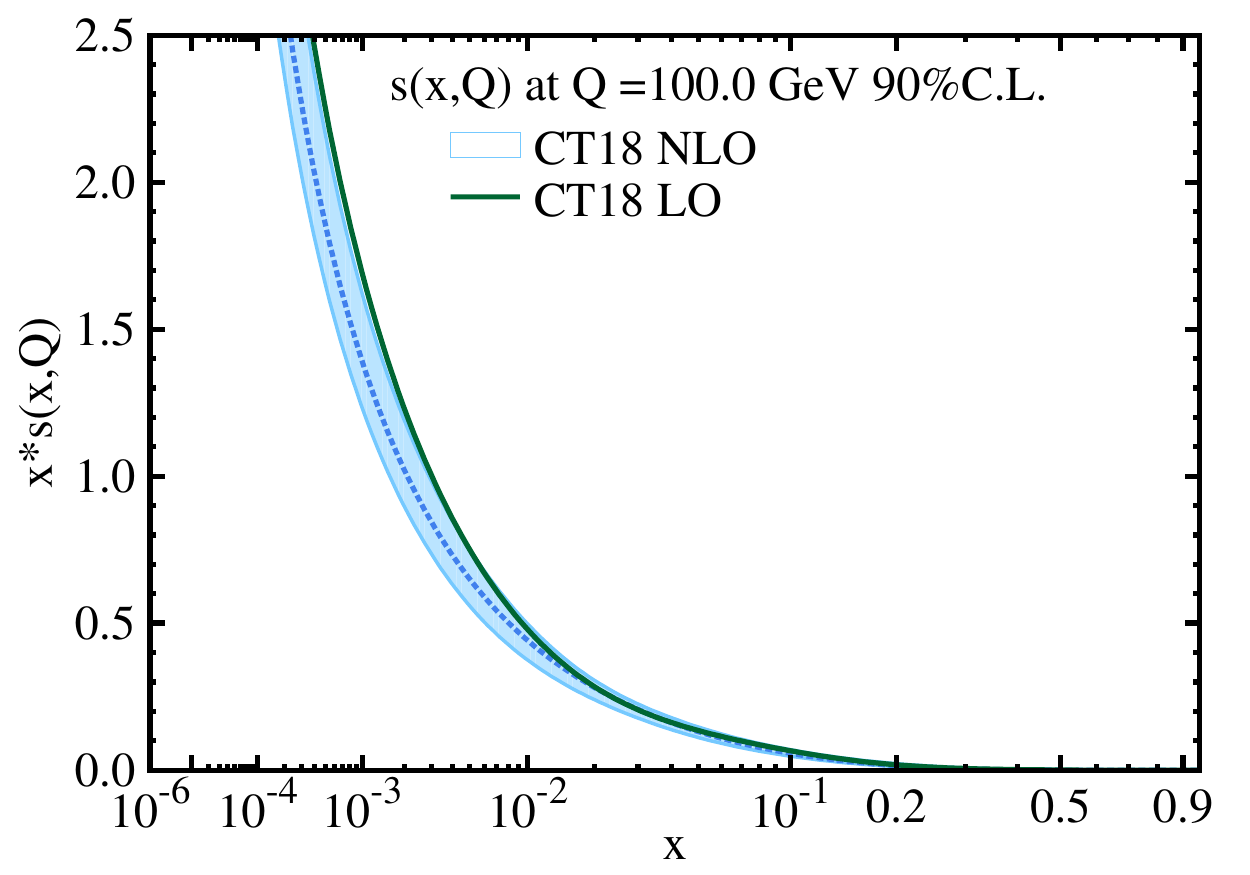}} 
\caption{The CT18 LO PDFs at $Q = 100$ GeV compared with CT18 NLO PDFs.}
\label{fig:PDF_configuration_100GeV} 
\end{figure}

The CT18 LO PDFs at 100 GeV are shown in Fig.~\ref{fig:PDF_configuration_100GeV}. For the $u$ and $d$ quarks distributions in the small-$x$ and large-$x$ limits, the CT18 LO is consistent with CT18 NLO up to its uncertainty bands. In Figs.~\ref{fig:PDF_configuration_100GeV:c} and~\ref{fig:PDF_configuration_100GeV:d}, the CT18 LO $u_v$ and $d_v$ distributions at 100 GeV are still different from the CT18 NLO distributions in the way similar to Fig.~\ref{fig:PDF_configuration_1.3GeV} at 1.3 GeV. The CT18 LO gluon PDF at 100 GeV is higher than CT18 NLO for $x < 3\times10^{-3}$. In Fig.~\ref{fig:PDF_configuration_100GeV:f}, the CT18 LO $s$ quark PDF at 100 GeV is consistent with CT18 NLO.

In Table~\ref{tab:2nd_moment}, we summarize the second moments $\langle x \rangle$ at the initial $Q_0$ scale, which quantifies the momentum carried by an individual flavour parton. 
Comparing to CT18 NLO central values, there are significant increments in the strange and gluon second moments. As mentioned before, more hard gluons are particularly required at the LO in order to describe the precision data, by increasing the parton densities of sea (anti)quarks and gluons in the smaller-$x$ region via the LO DGLAP evolution. Similarly, the parametrised strange PDF at $Q_0$ scale is also driven to acquire more momentum in order to describe data which are sensitive to $s$-quark PDF, such as DIS di-muon data and precision $W$ and $Z$ data. 
Due to the momentum sum rule, all parton densities are correlated. Hence, because of
the enhancements in ${\langle x \rangle}_s$ and ${\langle x \rangle}_g$, all the other flavours are allocated with less momenta than in CT18 NLO.
The CT18 LO second moments at $Q_0$ scale in general are consistent with CT18 NLO within one standard deviation, except for the strange PDF. Without higher-order corrections, a better determination of strange PDF is difficult, as to be seen below.

\begin{table}[h]  
\begin{center}  
\begin{tabular}{l|llll}  
PDFs & ${\langle x \rangle}_{u}$ & ${\langle x \rangle}_{d}$ &  ${\langle x \rangle}_{u_v}$ & ${\langle x \rangle}_{d_v}$ \\ \hline
CT18 LO & 0.3362 & 0.1571 & 0.3076 & 0.1201  \\
CT18 NLO & 0.3480(359) & 0.1692(402) & 0.3188(359) & 0.1317(401)  \\ \hline
 & ${\langle x \rangle}_{\bar{u}}$ & ${\langle x \rangle}_{\bar{d}}$ & ${\langle x \rangle}_{s}$ & ${\langle x \rangle}_{g}$ \\ \hline
CT18 LO & 0.0286 & 0.0370 & 0.0183 & 0.4045 \\
CT18 NLO & 0.0292(21) & 0.0375(26) & 0.0125(33) & 0.3911(101) \\ \hline
\end{tabular}  
\end{center}  
\caption{\label{tab:2nd_moment}
The second moments of CT18 LO and CT18 NLO PDFs at the initial scale $Q_0 = 1.3$ GeV.
All computed moment uncertainties are based on $68\%$ confidence level (CL).
}
\end{table}

\subsection{Impact of precise LHC $W$ and $Z$ production data}
\label{subsec:LHC_run2}

Today, vector boson production at the LHC can be measured so precisely that total experimental uncertainty is at a percent level. 
At NLO, the production cross-sections of the Drell-Yan processes receive large corrections from contributions of additional quark-gluon subprocesses and the virtual correction on the vertex of quark associated with vector boson,
so that the Born-level cross-section is not capable of describing experimental data as accurately as those beyond the leading order.
The fourth column of Table~\ref{tab:quality_of_fit} shows the values of $\chi^2/N_{\text{pt}}$ of the CT18 LOpert fit, which uses the LO theory prediction without including the $K$-factor, as introduced in Eq.~\eqref{Eq:K-fac}, for computing Drell-Yan cross-sections. It is evident that the CT18 LOpert fit cannot describe the data well, with a very large value of $\chi^2/N_{\text{pt}}$ for each individual Drell-Yan data. Furthermore, the resulting PDFs are also problematic, especially the strange quark PDF.
 
\begin{figure}[htbp]
\centering
\hspace{0.01\columnwidth} 
\centering \subfigure[$Q = 1.3$ GeV]{ \label{fig:ScanLHC2Weight:a} 
\includegraphics[width=0.475\columnwidth]{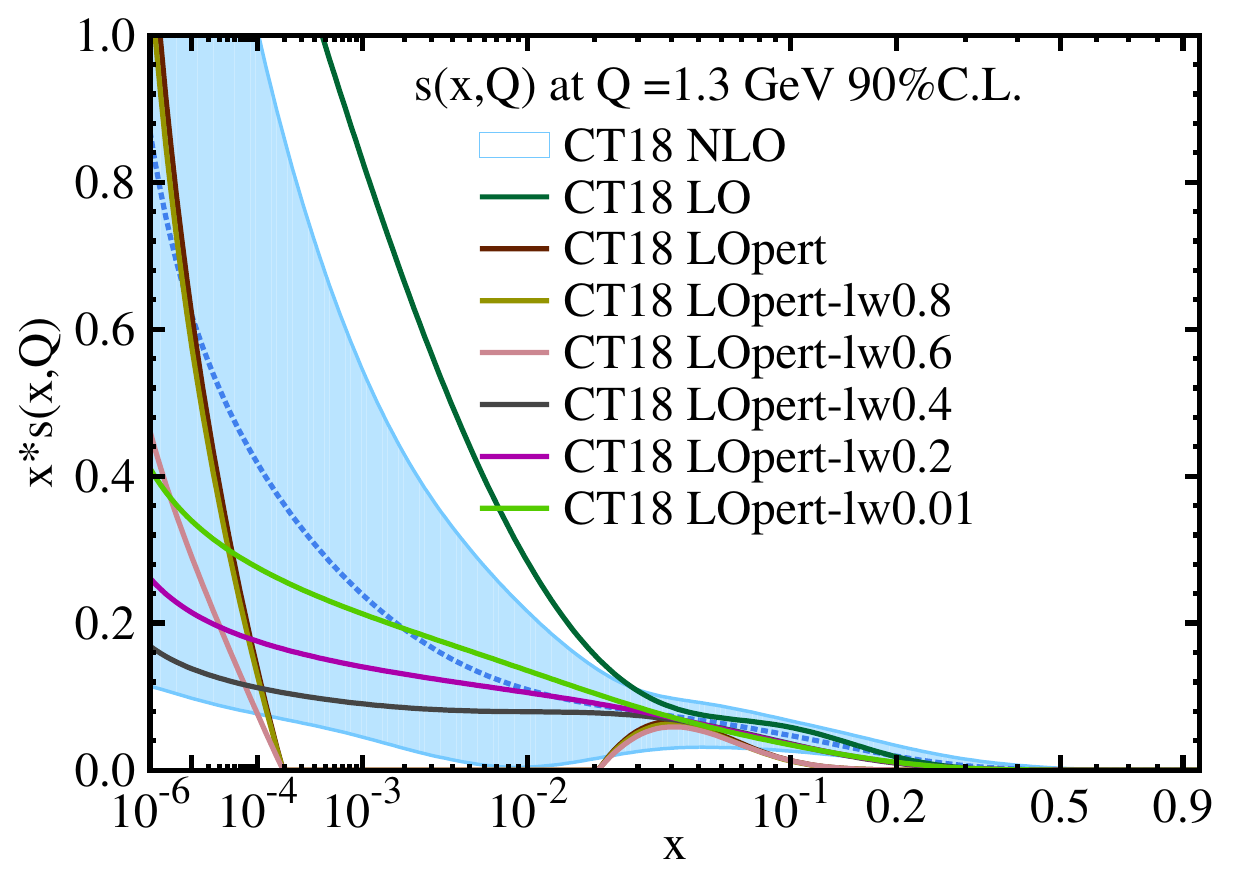}} 
\centering \subfigure[$Q = 100$ GeV]{ \label{fig:ScanLHC2Weight:b} 
\includegraphics[width=0.475\columnwidth]{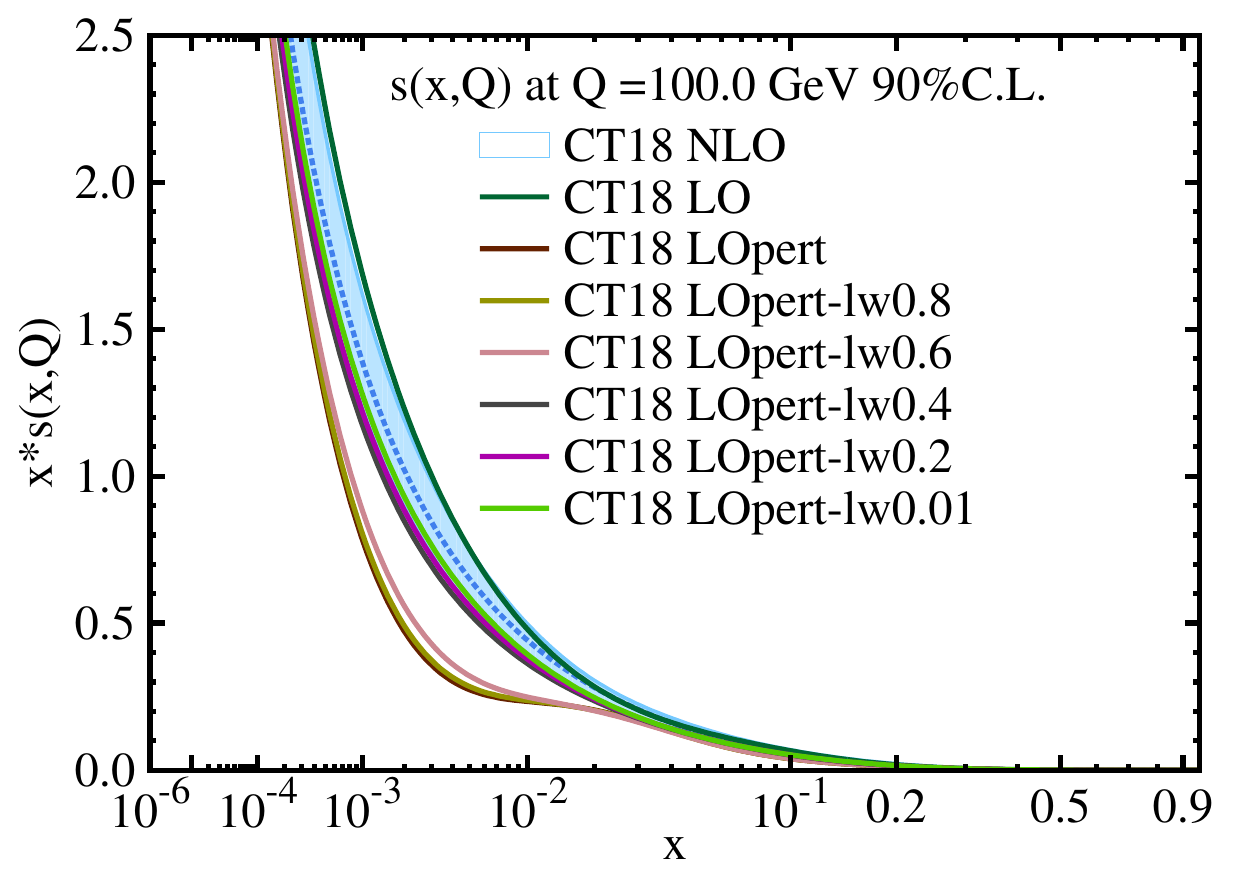}} 
\caption{The comparison of $s(x)$ at $Q_0 = 1.3$ GeV and 100 GeV among CT18 LO, 
CT18 LOpert series, and CT18 NLO. In CT18 LOpert series,
The suffix ``lw" means ``low weight", and the following numbers refer to the weights to ID 245, 246, 249, 250, 253. In the PDFs named ``CT18 LOpert", i.e., the brown curves in both panels, the weights to these data sets are unity, as in CT18 LO and CT18 NLO.
}
\label{fig:ScanLHC2Weight}  
\end{figure}

To illustrate the impact of these gauge boson production data sets, we compare a series of fits, 
starting from CT18 LOpert introduced in Sec.~\ref{subsec:treatment}, with various weights to LHC Run-II vector boson production data sets, namely, ID 245, 246, 249, 250, and 253, as shown in Fig.~\ref{fig:ScanLHC2Weight}.
For weights of these data sets being larger or equal to 0.6, the strange PDF vanishes in the range $2\times10^{-4} < x < 0.02$ at $Q_0$ = 1.3 GeV under the impact of these data sets. When evolved to 100 GeV,  the strange distribution is still quite small in this range of $x$. If the impact of these data sets is gradually removed from fits as weights becoming smaller, the resulting $s$-quark PDF will become larger. 
When fitting the vector boson production, the up and down sea quark distributions is driven by data to increase their magnitudes to compensate for the deficiency in the LO Wilson coefficients. Since all flavours are correlated under the total momentum sum rule, the magnitude of the strange PDF has to be reduced when that of the others is increased.
Such a strong suppression of the strange PDF is well resolved in CT18 LO by applying the Drell-Yan $K$-factor, Eq.~ \ref{Eq:K-fac}
to those Drell-Yan data sets, so that $\bar u$ and $\bar d$ PDFs are suppressed and $s$-quark PDF is increased as compared to those in CT18 LOpert.

\begin{figure}[htbp]
\centering
\hspace{0.01\columnwidth} 
\centering \subfigure[$W^+$ production]{ \label{fig:vbp_ID250:b} 
\includegraphics[width=0.475\columnwidth]{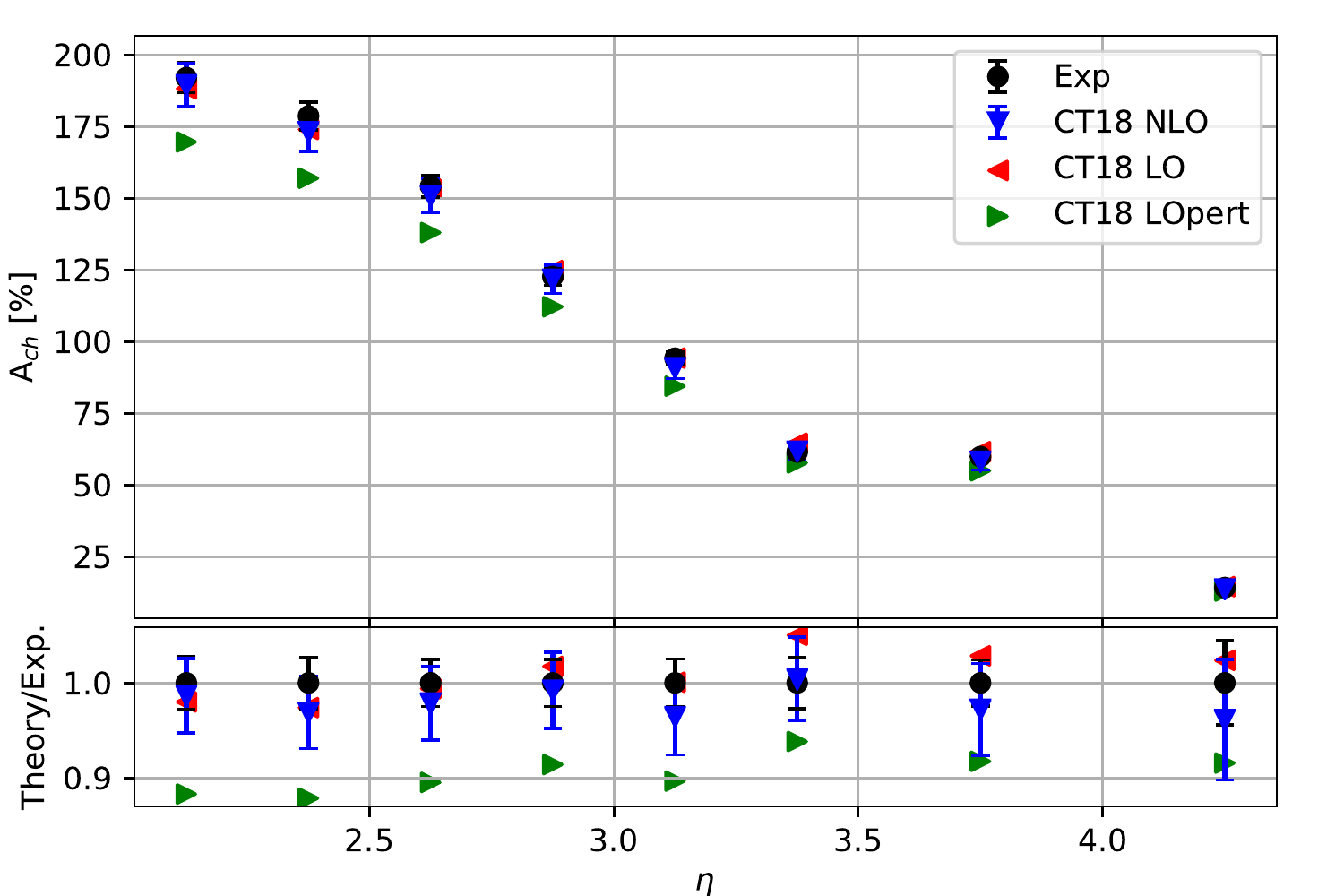}}
\centering \subfigure[$W^-$ production]{ \label{fig:vbp_ID250:a} 
\includegraphics[width=0.475\columnwidth]{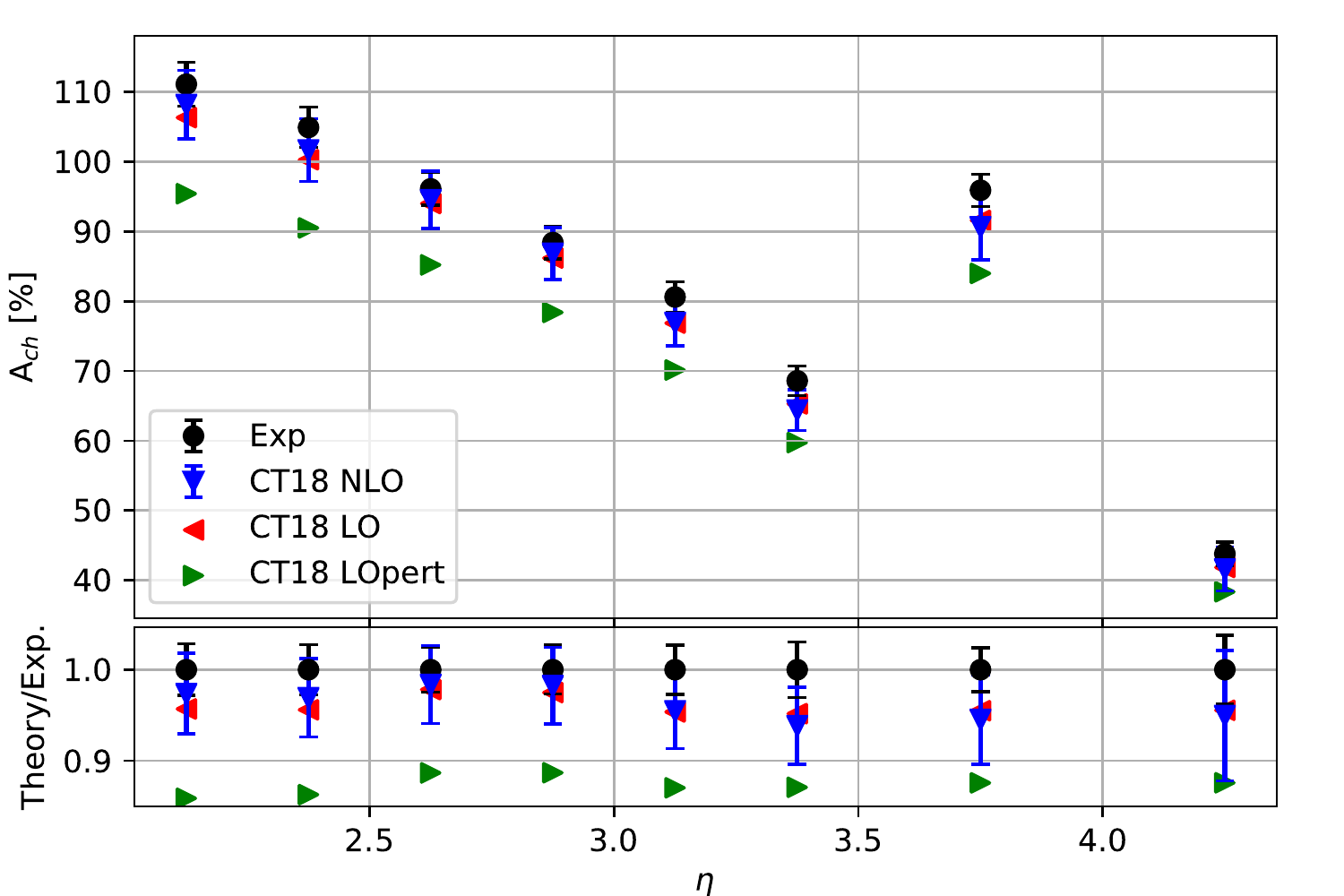}}
\centering \subfigure[$Z$ production]{ \label{fig:vbp_ID250:b} 
\includegraphics[width=0.475\columnwidth]{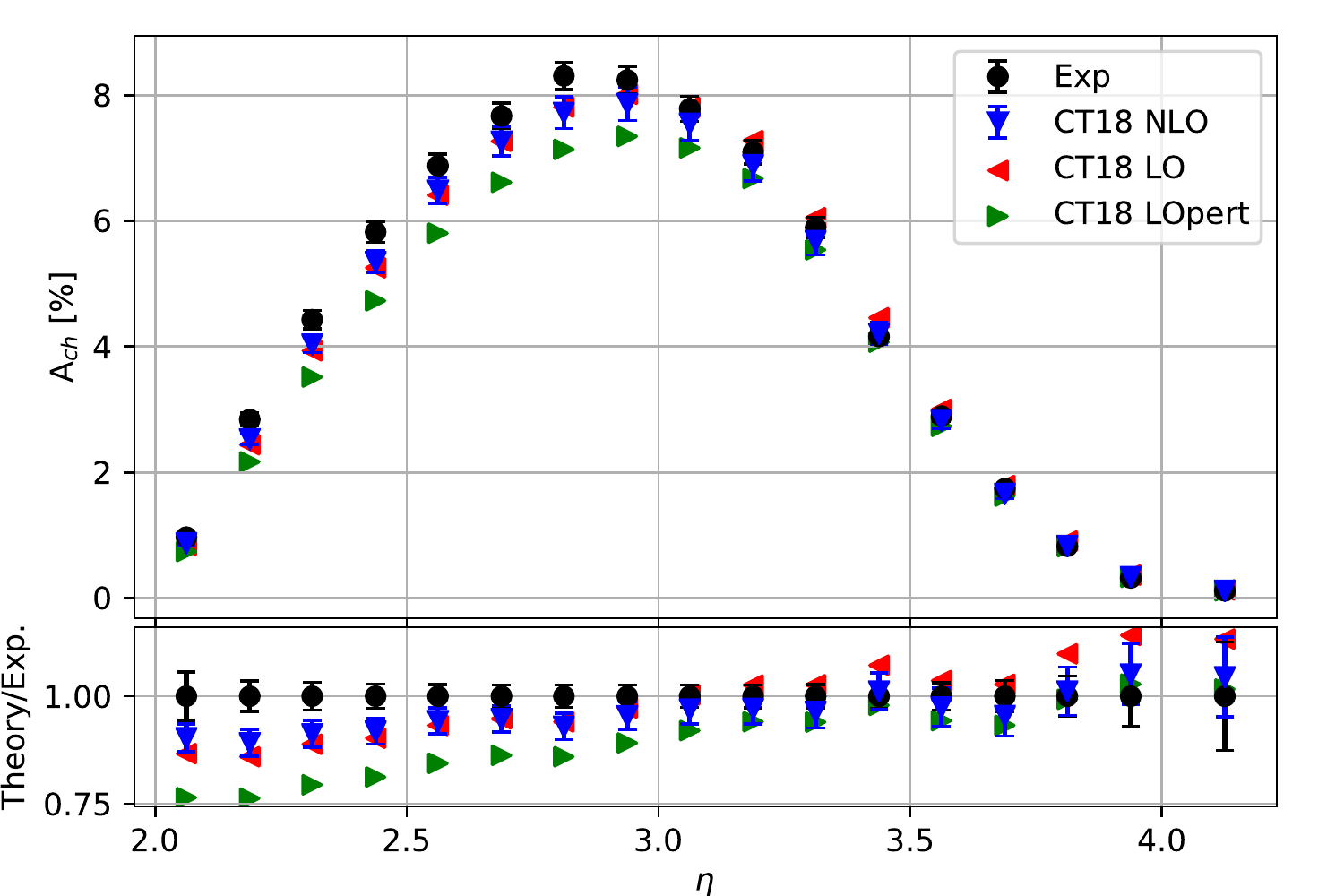}}
\caption{Predictions for the rapidity distribution of $W^{\pm}$ and $Z$ production at LHCb 7 TeV, ID 245, by CT18 LO, CT18 LOpert and CT18 NLO. Theory predictions are also compared to the experimental values.
}
\label{fig:vbp_ID245}
\end{figure}

As an example, in Fig.~\ref{fig:vbp_ID245}, 
we compare predictions for the ID 245 LHCb $W$ and $Z$ bosons production at 7 TeV~\cite{LHCb:2015okr} by CT18 LO, CT18 LOpert, and CT18 NLO 
to the experimental data points. The theory predictions are calculated by using the APPLgird package~\cite{Carli:2010rw}.
Without the application of $K$-factor, Eq. \ref{Eq:K-fac}, CT18 LOpert cannot provide enough cross-sections for either $W$ or $Z$ production, and yields a $\chi^2/N_{\text{pt}}$ as large as 8.36, as shown in Table~\ref{tab:quality_of_fit}. Such a difficulty in fitting vector bosons production data results in the vanishing feature of the CT18 LOpert $s$-quark PDF at lower energy scale as shown in Fig.~\ref{fig:ScanLHC2Weight}. In CT18 LO, the prediction of overall magnitude of this process, ID 245, is improved by the Drell-Yan $K$-factor and consistent with predictions by CT18 NLO. Hence, the quality of fit to this data is improved to $\chi^2/N_{\text{pt}} = 5.85$. Consequently, the suppression on the $s$-quark PDF as in CT18 LOpert is accordingly relaxed. But the improvement of the shape of rapidity spectrum still requires higher-order QCD corrections. 

\subsection{$\alpha_s$ dependence of CT18 LO}
\label{subsec:as}

The strong coupling constant is one of key elements in computing theory predictions, and hence fed as input into a PDFs global fit. For the PDFs fit beyond LO, it is widely accepted~\cite{Hou:2019efy, Ball:2021leu, Bailey:2020ooq, NNPDF:2017mvq, Dulat:2015mca, Harland-Lang:2014zoa, Martin:2009iq} that the value of strong coupling at $Z$-boson mass scale is fixed at its PDG value $\alpha_s(m_Z) = 0.118$~\cite{ParticleDataGroup:2020ssz}. 
Due to the missing of important quantum corrections,  to generate more sea quarks via parton evolution to resolve the difficulty in a LO PDF fit, the value of $\alpha_s(m_Z)$ as input is often fixed at a higher value than the PDG global average $\alpha_s(m_Z) = 0.118$.  A number of LO PDFs~\cite{Bailey:2020ooq, NNPDF:2017mvq, Dulat:2015mca} takes $\alpha_s(m_Z)$ to be at 0.130. For NNPDF4.0~\cite{Ball:2021leu}, NNPDF3.1~\cite{NNPDF:2017mvq}, and CT14~\cite{Dulat:2015mca}, LO PDFs with $\alpha_s(m_Z) = 0.118$ are also provided. For MSTW08 LO~\cite{Martin:2009iq} and MMHT14 LO~\cite{Harland-Lang:2014zoa} $\alpha_s(m_Z)$ is fixed at 0.140 and 0.135 respectively.

\begin{figure}[htbp]
    \centering
    \includegraphics[width=0.475\columnwidth]{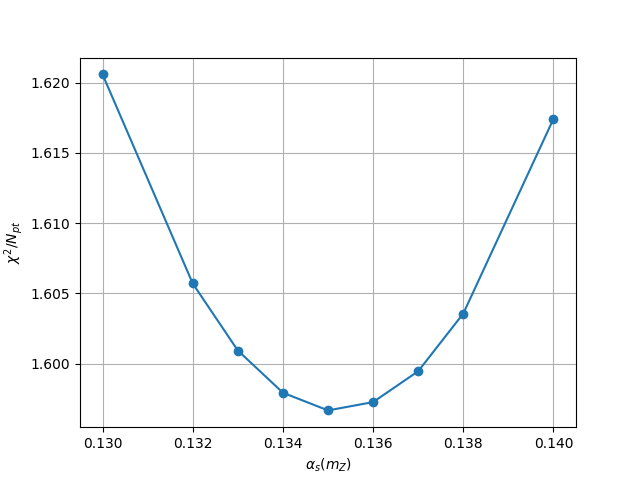}
    \caption{The $\chi^2/N_{\text{pt}}$ scan over various values of $\alpha_s(m_Z)$ for CT18 LO.}
    \label{fig:as_scan}
\end{figure}

Determination of the input value of $\alpha_s(m_Z)$ of CT18 LO PDFs is done by searching for the optimal value of this theoretical input parameter. We scan over various values of $\alpha_s(m_Z)$ by performing a series of fits with all the other theoretical and experimental setups identical to those of CT18 LO.
Fig.~\ref{fig:as_scan} shows the value of $\chi^2/N_{\text{pt}}$ of global fits while varying the value of $\alpha_s(m_Z)$.
As shown in the figure, the best-fit value of $\alpha_s(m_Z)$ is found to be around 0.135. Compared to the usual choice $\alpha_s(m_Z) = 0.130$ for most of the LO PDFs, a CT18 LO PDFs fit with $\alpha_s(m_Z) = 0.135$ could reduce the $\chi^2/N_{\text{pt}}$ by 0.024, equivalent to about 84 units for total $\chi^2$. A larger value of $\alpha_s(m_Z)$ than 0.135 cannot lead to a better fit to the data sets. 
Accordingly, we decide to fix the value of $\alpha_s(m_Z)$ to be 0.135 as input for the CT18 LO PDFs.

\section{Phenomenology}
\label{sec:pheno}

In this section, we present the implication of the CT18 LO PDFs by comparing some LHC phenomenologies generated with CT18 LO and NLO PDFs.
Specifically, for the differential distributions, we consider the experimental measurement of the charge asymmetry $A_{\text{ch}}$ in $W$-boson production at 8 TeV with the CMS detector. For the inclusive total cross-section, the prediction of the single-top production at 14 TeV is calculated. In these computations, the input physical parameters are set as follows: 

\begin{equation}
 m_W = 80.385 \, \text{GeV}, \quad G_F = 1.16639\times10^{-5} \,  \text{GeV}^{-2}, \quad \Gamma_W = 2.06 \, \text{GeV}.
 \label{eq:phys_para}
\end{equation}

\subsection{Charge asymmetry $A_{\text{ch}}$ in $W$-boson production}
\label{subsec:Ach}

\begin{figure}[htbp]
\centering
\hspace{0.01\columnwidth} 
\centering \subfigure[predictions to ID 249]{ \label{fig:Ach_ID249:a} 
\includegraphics[width=0.475\columnwidth]{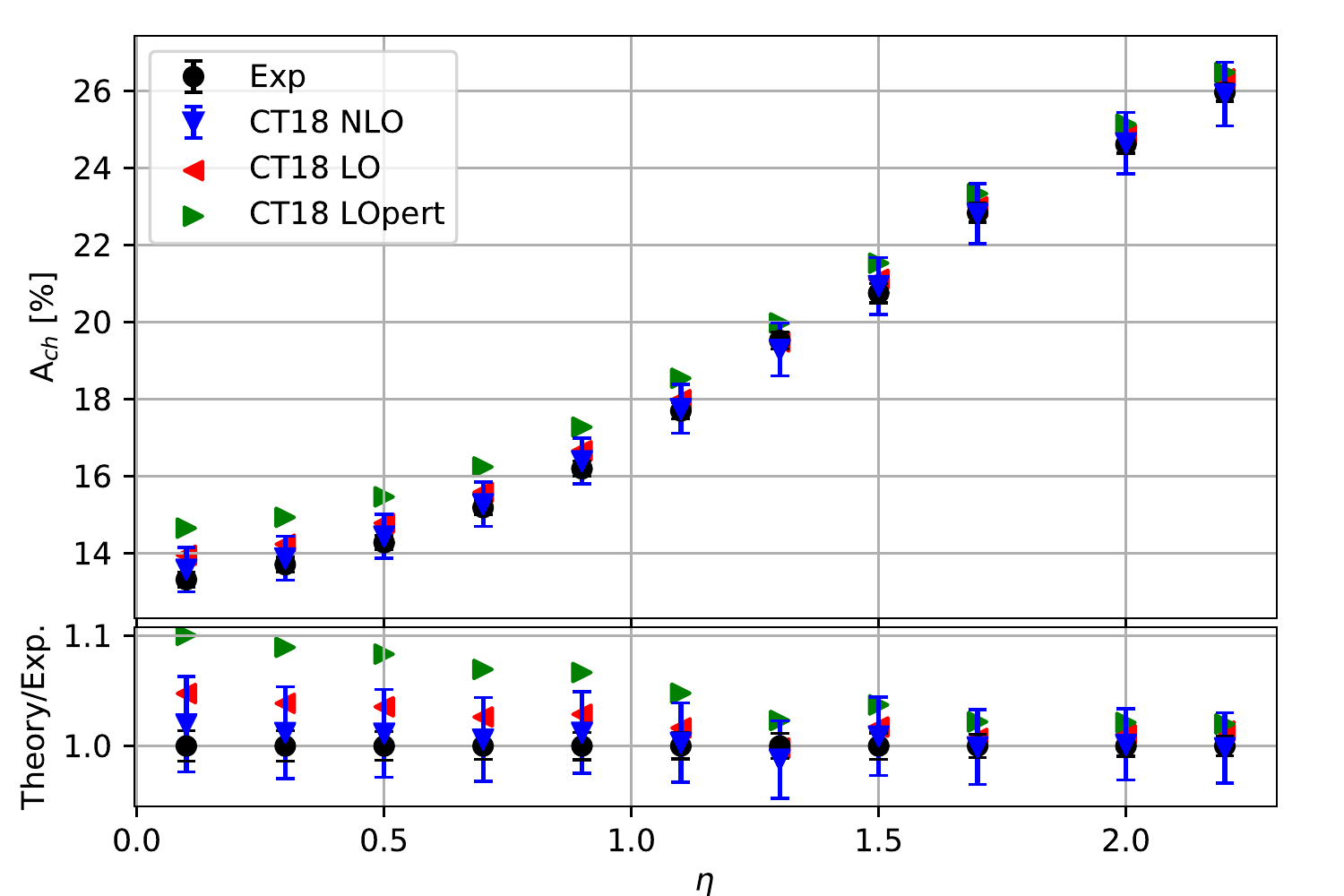}}
\centering \subfigure[ratios of $\bar{d}/\bar{u}$ at 100 GeV]{ \label{fig:Ach_ID249:b}  
\includegraphics[width=0.475\columnwidth]{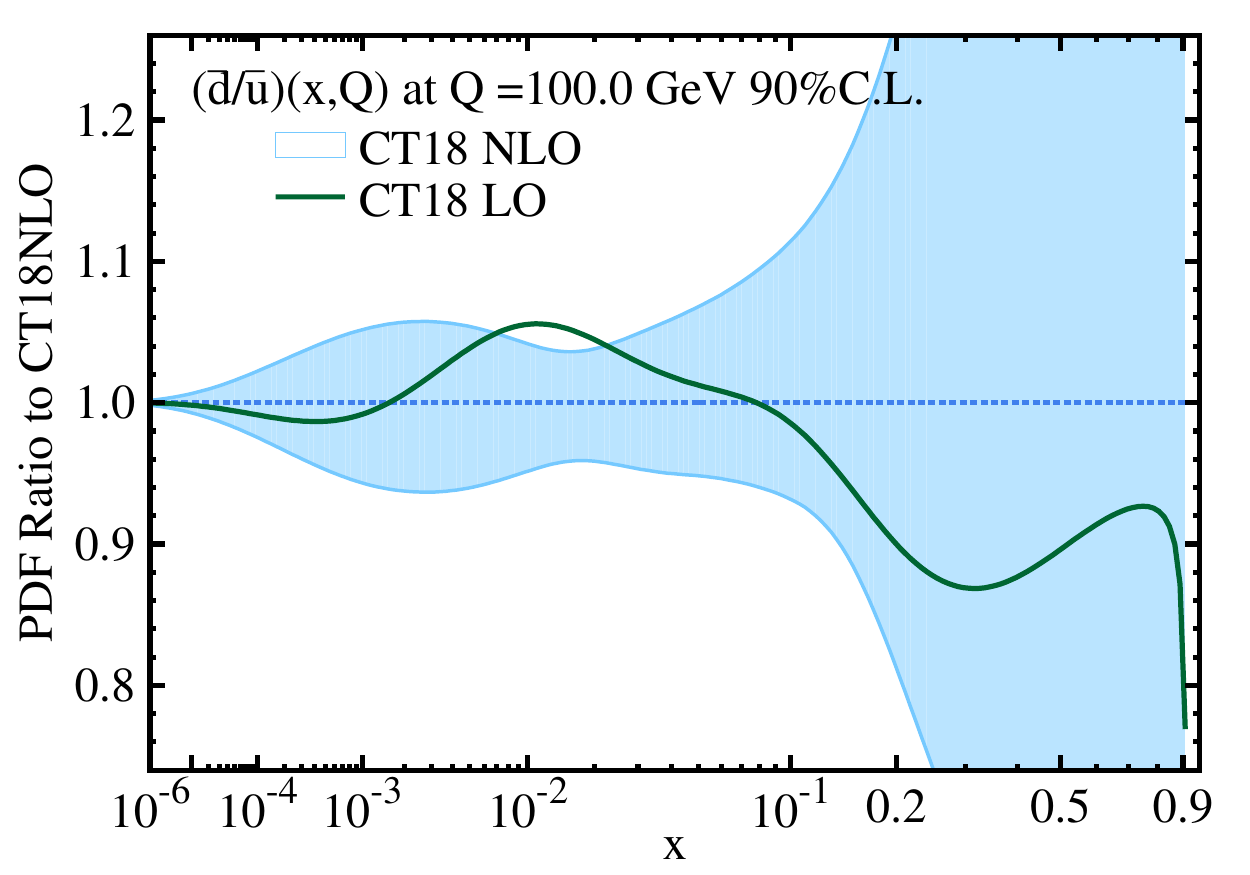}}
\caption{Left panel: similar to Fig.~\ref{fig:vbp_ID245}, but for CMS W-boson charge asymmetry $A_{\text{ch}}$, ID 249. Right Panel: the comparison of the ratio $\bar{d}/\bar{u}$ in between CT18 LO and CT18 NLO at 100 GeV.
}
\label{fig:Ach_ID249}
\end{figure}

The difference between $W^+$ and $W^-$ production cross-sections via the Drell-Yan process is dominated by PDFs of incoming quarks. 
Hence, the rapidity asymmetry of the charged lepton from $W$ boson decay
serves as a good observable to probe the ratio of parton luminosities. 
In Fig.~\ref{fig:Ach_ID249}, the comparison between predictions for ID 249 CMS 
muon charge asymmetry $A_{\text{ch}}$ at 8 TeV~\cite{CMS:2016qqr} and experimental data points are shown, as a function of the pseudo-rapidity of muon from $W$ boson decay. The calculation of $pp \to W^{\pm} +X \to \mu^\pm \nu +X$ differential cross-sections rapidity distributions are performed with APPLgrid~\cite{Carli:2010rw}.

As shown in the left panel of Fig.~\ref{fig:Ach_ID249}, predictions by CT18 LO exhibits a different shape from the experimental data. For lower rapidity region, the CT18 LO predictions are about $\sim$5\% higher than the measurements. For larger rapidity, the CT18 LO predictions for $A_{\text{ch}}$ tends to be more consistent with measurements. While the CT18 LO predictions for muon charge asymmetry generally lies within the CT18 NLO uncertainty band over whole rapidity range, the CT18 LOpert prediction shows a worse agreement. This phenomenon indicates that the CT18 LO and NLO are consistent in the ratio of down and up antiquark PDFs, which can be directly observed in the right panel of Fig.~\ref{fig:Ach_ID249}, where the CT18 LO is mostly inside of CT18 NLO error band for the ratio of $\bar{d}/\bar{u}$, except for the range $0.01 < x < 0.03$ where the CT18 LO $\bar{d}/\bar{u}$ is outside of CT18 NLO error band.

\subsection{Single-top production}

We select the calculation of total cross-section for the $t$-channel inclusive single-top production as a representative process to study the implication of CT18 LO PDFs.

This process plays an important role in constraining the heavy quark PDF, and it has been measured at LHC~\cite{ATLAS:2012byx,ATLAS:2014sxe,CMS:2011oen,CMS:2012xhh,ATLAS:2017rso,CMS:2014mgj,ATLAS:2016qhd,CMS:2016lel,CMS:2018lgn} at various center-of-mass energies.
The theoretical calculation of this process could serve as a test on the consistency in PDFs at different perturbation orders~\cite{Sullivan:2017aiz, Campbell:2021qgd}, since the total inclusive cross-sections consistently predicted at different orders are all expected to reproduce the data.
We make use of this property to illustrate the consistency of CT18 LO with CT18 NLO.
In our calculation, the $t$-channel inclusive single-top production cross-section is computed
by MCFM~\cite{Campbell:2010ff, Boughezal:2016wmq, Campbell:2015qma}. For this calculation, the input parameters take the values as shown in Eq. \ref{eq:phys_para}, along with the renormalization and factorization scales chosen as $\mu_R = \mu_F = m_t$,
while the top mass $m_t$ is chosen to be consistent with the corresponding PDF sets.

Predictions for the $t$-channel inclusive single-top production with a variety of PDFs are presented in Table~\ref{tab:single_top}. In general, due to the lack of higher-order corrections, the LO predictions for this process tend to be smaller than their corresponding NLO predictions. The CT18 LO prediction of single top quark production is slightly outside of the CT18 NLO uncertainty band, while for the single anti-top quark production the CT18 LO and CT18 NLO are well consistent.
Comparing to CT14 LO predictions, the CT18 LO predictions to both top and anti-top production are enhanced substantially,  
and better consistent with its corresponding NLO fit.

\begin{table}[htbp]
\begin{center}  
\begin{tabular}{ll|ll|ll}  
PDFs & $\alpha_s(m_Z)$ & $\sigma_{LO}(t)$ [pb] & $\sigma_{LO}(\bar{t})$ [pb] & $\sigma_{NLO}(t)$ [pb] & $\sigma_{NLO}(\bar{t})$ [pb] \\ \hline \hline
CT18 LO & 0.135 & 153.5 & 93.5 & - & - \\ 
CT18 NLO~\cite{Hou:2019efy} & 0.118 & - & - & 156.8 $\pm$ 2.5 & 93.4 $\pm$ 1.6 \\ \hline
CT14 LO~\cite{Dulat:2015mca} & 0.118 & 137.6 & 80.6 & - & - \\
CT14 NLO~\cite{Dulat:2015mca} & 0.118 & - & - & 157.0 $\pm$ 3.3 & 93.2 $\pm$ 1.9 \\ \hline
MSHT20 LO~\cite{Bailey:2020ooq} & 0.130 & 142.3 $\pm$ 1.0 & 97.3 $\pm$ 0.6 & - & - \\
MSHT20 NLO~\cite{Bailey:2020ooq} & 0.118 & - & - & 156.5 $\pm$ 1.1 & 94.7 $\pm$ 0.7 \\ \hline
NNPDF4.0 LO~\cite{Ball:2021leu} & 0.118 & 133.6 $\pm$ 1.1 & 83.2 $\pm$ 0.8 & - & - \\
NNPDF4.0 NLO~\cite{Ball:2021leu} & 0.118 & - & - & 155.1 $\pm$ 0.9 & 93.7 $\pm$ 0.5 \\ \hline
NNPDF3.1 LO~\cite{NNPDF:2017mvq} & 0.118 & 141.7 $\pm$ 1.8 & 85.4 $\pm$ 2.0 & - & - \\
NNPDF3.1 LO~\cite{NNPDF:2017mvq} & 0.130 & 153.5 $\pm$ 1.8 & 92.6 $\pm$ 1.6 & - & - \\
NNPDF3.1 NLO~\cite{NNPDF:2017mvq} & 0.118 & - & - & 155.2 $\pm$ 1.2 & 93.4 $\pm$ 0.7 \\ \hline
NNPDF3.0 LO~\cite{NNPDF:2014otw} & 0.118 & 149.1 $\pm$ 13.1 & 90.5 $\pm$ 8.5 & - & - \\
NNPDF3.0 LO~\cite{NNPDF:2014otw} & 0.130 & 159.4 $\pm$ 7.7 & 96.6 $\pm$ 5.1 & - & - \\
NNPDF3.0 NLO~\cite{NNPDF:2014otw} & 0.118 & - & - & 162.6 $\pm$ 1.9 & 99.2 $\pm$ 1.5 \\ \hline
HERAPDF20 LO~\cite{H1:2015ubc} & 0.130 & 154.6 $\pm$ 0.8 & 92.1 $\pm$ 0.7 & - & - \\
HERAPDF20 NLO~\cite{H1:2015ubc} & 0.118 & - & - & 162.5 $\pm$ 0.9 & 96.5 $\pm$ 0.7 \\ \hline
\end{tabular}  
\end{center}  
\caption{\label{tab:single_top} The total cross-sections of inclusive single-top production in $pp$ collision at 14 TeV at LO and NLO for various PDFs sets.
}
\end{table}

\section{Conclusion}
\label{sec:conclusion}

In this paper, we present CT18 LO PDFs, which is obtained within the general framework of CT18 global analysis with extensions of two special treatments, as defined in Sec.~\ref{subsec:treatment}. 
One is to discard some data sets, which cannot be properly described at LO (such as Drell-Yan data with different cuts on the transverse momemta of the two final state leptons), from the CT18 data set. The other is to apply a $K$-factor to predictions for Drell-Yan processes for making up the insufficiency of LO predictions.
As the result, the quality of the LO fit, cf. CT18 LO, is substantially improved from that of a naive LO fit, cf. CT18 LOpert. 
In CT18 LOpert, strange quark distributions are strongly impacted by the high-precision $W$ and $Z$ production data from LHC Run-II era, as shown in Fig.~\ref{fig:ScanLHC2Weight}, since LO Wilson coefficients cannot provide enough normalization to predictions to these precise measurements. The strong suppression on $s(x)$ is relaxed in CT18 LO via the implementation of the Drell-Yan $K$-factor, which supplies additional normalization to Drell-Yan processes.
We have checked that CT18 LO PDFs is capable in generating numerical predictions close to CT18 NLO PDFs for the rapidity distribution of charge asymmetry in $W$-boson production and the total cross-section of $t$-channel inclusive single-top production. 
But still we should stress that the CT18 LO PDFs is different from CT18 NLO PDFs on many aspects, including the quality of fit, PDFs configurations, and the ability of describing experimental data. Therefore we would not suggest to use this result, CT18 LO, in analyses where precision is the dominant requirement. Since the LO PDF fits embed a huge theoretical uncertainty, we do not provide an error set for the CT18 LO PDFs.

The central CT18 LO PDFs in LHAPDF format~\cite{Buckley:2014ana} is publicly available:
\begin{center}
\url{https://lhapdf.hepforge.org}     
\end{center}

\subsection*{Acknowledgements}
PMN is partially supported by the U.S. Department of
Energy under Grant No. DE-SC0010129. 
The work of CPY is partially supported by the U.S. National Science Foundation under Grant No. PHY-
2013791. CPY is also grateful for the support from the Wu-Ki Tung endowed chair in particle physics.

%

\bibliographystyle{utphys}
\bibliography{refs}
\end{document}